\documentclass[a4paper,12pt]{article}

\usepackage[scale={.75,.8}]{geometry} 
\usepackage{pstricks}
\usepackage[dvips]{graphicx}
\usepackage[latin2]{inputenc}
\usepackage{epic}
\usepackage{latexsym}
\usepackage{epsf}
\usepackage{epsfig}
\usepackage{amssymb,amsmath}

\newcommand{\eref}[1]{eq.\ (\ref{e.#1})} 
\newcommand{\erefn}[1]{ (\ref{e.#1})}
 
\newcommand{\aref}[1]{\ref{a.#1}}

\newcommand{\cref}[1]{Chapter \ref{c.#1}}

\def\nn{\nonumber \\}  
\newcommand{\nl}{& \nonumber \\ &}
\newcommand{\bnl}{\right . & \nonumber \\ & \left .}

\def\ds{\displaystyle}

\def\beq{\begin{equation}} 
\def\eeq{\end{equation}} 
\newcommand{\ba}{\begin{array}}  
\newcommand{\ea}{\end{array}} 
\newcommand{\bea}{\begin{eqnarray}}  
\newcommand{\eea}{\end{eqnarray} }  
\newcommand{\bal}{\begin{align}}
\newcommand{\eal}{\end{align}}   
\def\bi{\begin{itemize}}  
\def\ei{\end{itemize}}  
\def\ben{\begin{enumerate}}  
\def\een{\end{enumerate}}  
\def\beq{\begin{equation}}  
\def\eeq{\end{equation}}  
\def\bc{\begin{center}}
\def\ec{\end{center}} 
 \def\bt{\begin{table}}
\def\et{\end{table}}  
 \def\btb{\begin{tabular}}
\def\etb{\end{tabular}}

\def\cl{{\mathcal L}}  
   
\def\cm{{\mathcal M}}  
\def\co{{\mathcal O}}   
   
\def\cv{{\mathcal V}} 

\def\tev{\, {\rm TeV}}

\def\mass2{mass${}^2$}

\newcommand{\mpl}{M_{\mathrm{Pl}}}  
  
\newcommand{\mkk}{M_{\mathrm{KK}}}  
\def\ads{$\bf \mathrm{AdS}_5$}

\def\pa{\partial}

\newcommand{\tr}{\mathrm T \mathrm r}  

\def\rt{\sqrt{2}} 
\def\simlt{\stackrel{<}{{}_\sim}}

\def\ra{\rangle}
\def\la{\langle}

\newcommand{\ha}{{\hat a}}

\newcommand{\ti}{\tilde}  
\def\hc{{\rm h.c.}}

\begin{document}
  \begin{titlepage}
    \bigskip{} {\par\centering \textbf{
    Modelling strong interactions and longitudinally polarized vector boson scattering} \large \par}
    \bigskip{}
    
	    {\par\centering Adam Falkowski$^{1,2}$, Stefan Pokorski$^{2}$, 
	      J. P. Roberts$^{2}$\\ \par}
	    \bigskip{}
	    
		    {\par\centering
		      {
			{\small $^1$ CERN Theory Division, CH-1211 Geneva 23, Switzerland
			}\\
			{\small $^2$ Institute of Theoretical Physics, Warsaw University, 
			  \\ Ho\.za 69, 00-681 Warsaw, Poland}
			\par}
		    }
		    
		    \bigskip{}

		    \begin{abstract}
		      \noindent
We study scattering of the electroweak gauge bosons in 5D warped models. 
Within two different models we determine the precise  manner in which the Higgs boson and the vector resonances ensure the unitarity of longitudinal vector boson scattering. 
We identify three separate scales that  determine the dynamics of the scattering process in all cases. 
For a quite general background geometry of 5D, these scales can be linked to a simple functional of the warp factor. 
The models smoothly interpolate between a `composite' Higgs limit and a Higgsless limit. 
By holographic arguments, these models provide an effective description of vector boson scattering in 4D  models with a strongly coupled electroweak breaking sector.

		    \end{abstract}
\end{titlepage}

\newpage

\setcounter{footnote}{0}

\renewcommand{\theequation}{\arabic{section}.\arabic{equation}} 

\section{Introduction}
\label{sec:intro}

The mechanism of electroweak symmetry breaking is still unknown.
Within the SM, electroweak symmetry is broken by the condensation of a
weakly coupled elementary scalar field, the Higgs field. This simple
mechanism is consistent with electroweak precision measurements if the
mass of the Higgs boson is within the range $100-200~$GeV. However any
such fundamental scalar that is much lighter than the SM cut-off is
unnatural as there is no symmetry protecting its mass.

There are two primary approaches to solving this hierarchy problem of
the SM. The first, supersymmetry, provides a rationale for elementary
scalars and protects the Higgs boson masses from large quantum
corrections. The simplest implementation - the MSSM - ensures
perturbative physics up to the Planck scale and provides several
interesting predictions at the TeV scale.  Unfortunately it is not
free from some residual tuning of parameters once experimental
constraints are imposed.

A radically different idea is that a new, strongly interacting, sector
provides a TeV scale cut-off to the SM.  One can envisage Higgsless
electroweak symmetry breaking that is generated in a manner similar to
the chiral symmetry breaking in QCD.  The longitudinal components of
the $W$ and $Z$ bosons are provided by three Goldstone bosons of the
strongly interacting sector.  However breaking electroweak symmetry
with a strongly interacting sector does not necessarily lead to a
Higgsless theory.  It is also possible to construct models in which
the full Higgs doublet emerges as a composite particle.  An
interesting subset of such models are those in which the composite
Higgs doublet arises as a pseudo-Goldstone boson of some spontaneously
broken approximate global symmetry of the strongly interacting theory.

Strongly interacting theories are notoriously difficult to handle in
four dimensions.  It has been suggested \cite{holo}, however, that they may have a
`holographic dual' description in terms of a 5D gauge theory in a
warped background \cite{rasu}.  Modelling strong interactions by 5D
theories has become a useful tool, allowing for quantitative studies
of both QCD \cite{qcd} and electroweak symmetry breaking \cite{ew,agcopo}.

It is well known \cite{chga}, and has been recently emphasised in \cite{gigrpo,chchco}
that scattering of longitudinally polarized $W$ and $Z$ bosons may be
used as a probe of the dynamics that breaks electroweak symmetry.
Therefore in this paper we use the calculation of the $W$ and $Z$
boson scattering amplitudes to analyse and compare different 5D
descriptions of the electroweak symmetry breaking sector.

It is interesting to systematize the phenomenology of this holographic approach.
In this paper we extract the common features that show up in in gauge boson scattering that are
independent of such details of the model building as the symmetries of
the strongly interacting sector or the warp factor describing the 5D
geometry.
The recurring feature  is the appearance of three distinct physical scales:
\bi
\item $v$: the electroweak breaking scale that sets the mass of W and Z.
\item $f_h$: the scale that sets the compositeness scale of the Higgs, referred to as the Higgs decay constant. 
\item $\mkk$: the resonance scale that sets the mass of the first resonance .
\ei 
How these scales emerge from the 5D dual is a model dependent question. 
However, we show that the separation between these scales does not depend on the fine-grained details of the model.  
More precisely one can define a simple functional, which we call the volume factor, that depends on the size and the geometry of the 5th dimension. 
The volume factor fixes the ratio $f_h/\mkk$ and,  in the Higgsless case, also $v/\mkk$.  
The same volume factor also fixes  $\Lambda/\mkk$ where $\Lambda$ is the strong coupling scale at which the 5D effective description breaks down.

We discuss in some detail how the aforementioned scales show up in the gauge boson scattering amplitudes.
In the Standard Model the exchange of a Higgs boson cancels the divergent behaviour of the four point gauge boson vertex.
This cancellation is not maintained in more complex models of electroweak symmetry breaking. 
As can be expected, the violation of unitarity is associated with the scale $f_h$ that controls departures of the Higgs couplings from the Standard Model, while the full restoration of unitarity is postponed until the
resonance scale $\mkk$. 
We present quantitative results for the scattering amplitudes in two different 5D models.  
One is the model or ref. \cite{agdema} describing a composite Higgs emerging from a strong sector with the $SO(4)$ custodial symmetry. 
The other is the model of ref. \cite{agcopo} describing a pseudo-Goldstone Higgs from breaking 
$SO(5) \to SO(4)$ by strong interactions.  

We also consider the Higgsless limit of the 5D models.  This is the
limit where the Higgs boson decouples from the electroweak bosons and
plays no role in restoring unitarity, even though it may remain in the
physical spectrum.  In this case the electroweak scale $v$ becomes
intimately tied to the geometry of the 5th dimension and equals $f_h$.
The Higgsless limit turns out to be particularly insensitive to the
details of 5D modelling.

The paper is structured as follows. In section \ref{sec:SM} we review
the unitarisation of the gauge boson scattering amplitudes in the SM.
We employ the equivalence theorem that allows us to calculate the
scattering in terms of scattering of the Goldstone bosons eaten by $W$
and $Z$.  This serves to highlight the role of the Higgs boson in the
unitarisation and to fix our notation for the rest of the paper. 
 In section \ref{sec:strongParam} we discuss in general terms the manner in which 
strongly coupled electroweak sectors affect the longitudinal vector boson scattering.
In section \ref{sec:strongHolo} we turn to modelling a strongly interacting electroweak breaking
sector using a 5D holographic dual.  We investigate the 5D model
proposed in \cite{agdema} with the Higgs sector localized on the IR
brane and custodial symmetry in the bulk.  We calculate the couplings
of the Goldstone bosons to the physical Higgs and to the resonances,
and employ useful approximations that reveal a simple pattern in these
couplings.  In section \ref{s.hpgh}, we repeat this program for a 5D
model of the pseudo-Goldstone Higgs \cite{agcopo}.  In section
\ref{s.d} we collect the results of our quantitative studies and use
them to calculate the precise form of the WZ scattering amplitude. 
We present our conclusions in section \ref{sec:conc}.
Three appendices contain more technical details of our computations.  

\section{Gauge boson scattering in the Standard Model}
\label{sec:SM}
\setcounter{equation}{0} 

First we review the unitarisation of longitudinal gauge boson
scattering in the SM.  Here we use the equivalence theorem (ET) to
calculate the scattering amplitudes via the Goldstone bosons
\cite{et}.  This serves to fix our notation, and to highlight the role
of the Higgs boson in unitarising the amplitudes.

The Lagrangian for the Higgs doublet is:
\begin{equation}
\mathcal{L}=|\partial_\mu H|^2 -V(H^\dagger H) \label{hlagr}
\end{equation}

We parameterise the Higgs fields non-linearly:
\begin{eqnarray}
\label{e.smhnlp}
H&=&\frac{1}{\sqrt{2}} (v + h) U\left(\begin{array}{c} 0\\
1\end{array}\right),\\ \nonumber \text
{where} &&
U = e^{\left(\frac{iG_a\sigma_a}{v}\right)}=\cos\left(\frac{G}{v}\right)
+i\frac{G_a\sigma_a}{G}\sin\left(\frac{G}{v}\right),
\quad G^2=G_a G_a
\end{eqnarray}
where $h$ is the physical Higgs boson, $v$ is the Higgs vev and $G_a$
are the three Goldstone bosons.
Inserting this into our Lagrangian (\ref{hlagr}) we get:
\begin{eqnarray}
\nonumber\mathcal{L}&=&\frac{1}{2}\left(\partial_\mu h\right)^2-V(h)\\
&+&
{1 \over 2} \left(1+\frac{h}{v}\right)^2\left[\left(\partial_\mu
G\right)^2 + \frac{\sin^2(G/v)}{(G/v)^2}\left(\left(\partial_\mu
G_a\right)^2-\left(\partial_\mu
G\right)^2\right)\right]
\end{eqnarray}

From this we acquire canonically normalized kinetic terms and the
interaction terms of the Goldstone bosons and the Higgs boson that we
need to calculate the scattering.  The relevant terms are:
\begin{eqnarray}
\mathcal{L}_{G^4}&=&\frac{1}{6v^2}\left(\left(G_a\partial_\mu G_a
\right)^2 - \left(\partial_\mu G_a\right)^2
G_bG_b\right)\label{eq:G^4SM}\\
\mathcal{L}_{G^2h}&=&
\frac{h}{v}\left(\partial_\mu G_a\right)^2
\label{eq:G^2hSM}
\end{eqnarray}

To complete the picture we introduce the $SU(2)_L \times U(1)_Y$ gauge fields. 
 The $W^\pm$ bosons acquire longitudinal polarizations by eating the Goldstone
modes $G^\pm=(G_1\mp iG_2)/\sqrt{2}$ while the Z boson eats $G_3$.
Moreover, there appear three-point vertices involving gauge bosons:
\begin{eqnarray}
\nonumber \mathcal{L}=&-&i( G^-\partial_\mu G^+ - G^+\partial_\mu
G^-)(g_\gamma A_\mu + g_Z Z_\mu)\\ \nonumber &-& i\left(G_3\partial_\mu G^- -
G^-\partial_\mu G_3\right)g_WW_\mu^+\\ &-& i\left(G_3\partial_\mu G^+
- G^+\partial_\mu G_3\right)g_WW_\mu^-
\end{eqnarray}
where $g_\gamma=e,~g_Z=(g_L^2-g_Y^2)/2\sqrt{g_L^2+g_Y^2}$, 
$g_W = g_L/2$ and  $g_L$, $g_Y$ are the SM gauge couplings. 

The Higgs boson plays a crucial role in unitarising scattering
processes in which the initial and final state particles are $W$ or
$Z$. Using the ET we can calculate the leading order contribution to the  relevant scattering
processes using the following amplitudes for Goldstone boson
scattering. 
Here we take the process $W_L Z_L \to W_L Z_L$ as an example:
\begin{eqnarray}
 & \ds
 \cm_{G^+ G^3 \to G^+ G^3} =   \frac{t}{v^2}  -   \frac{t}{v^2} \frac{t}{t - m_h^2} 
\nl  \ds
-  g_W^2  \left (\frac{t - s}{u - m_W^2} + \frac{t - u}{ s - m_W^2} \right )  
\end{eqnarray}
Via the ET, this amplitude corresponds to the amplitude for $W_L^\pm
Z_L \rightarrow W_L^\pm Z_L$, up to terms $\mathcal{O} (m_W/E)$.  The
first term grows quadratically with energy.  This leads to unitarity
violation at high energies, unless it is cancelled by the term from
Higgs boson exchange that follows.  Therefore the presence of a
sufficiently light Higgs boson restores unitarity in the theory.  The
last term from the $W$ boson exchange is irrelevant to the discussion
of quadratic divergences as it contributes no growing term, but we
have included it for later convenience.

\section{Parameterising strongly coupled electroweak sectors}
\label{sec:strongParam}
\setcounter{equation}{0} 

We now go on to examine the question of the unitarity of gauge boson
scattering in theories  with an extended  electroweak sector.
In the SM unitarisation of the gauge boson scattering amplitude relies on the cancellation
between the quartic Goldstone vertex and the Higgs exchange diagrams.
This requires a precise correlation between the quartic Goldstone
self-coupling and the Higgs-Goldstone coupling.  In the following we
will discuss 5D models of electroweak symmetry breaking and
investigate how they affect this correlation. 

Since 5D warped physics provides a holographic description of strongly
coupled theories, we expect that deviations from the SM scattering
amplitudes will depend directly upon the compositeness scale of the
Higgs boson, which we denote as $f_h$. %
The composite structure should reveal itself in modifications of Higgs-Goldstone couplings by terms of
order $1/f_h^2$. 
 We thus expect the couplings to have the form:
\begin{equation}
\cl_{G^2 h} = g_h  \frac{h}{v}  (\pa_\mu G^a)^2   
\end{equation} 
where $g_h = 1 - \co (v^2/f_h^2)$.  For $g_h < 1$ the Higgs boson on
its own cannot unitarise the scattering of the gauge bosons:
\beq
\cm_{G^+G^3\rightarrow G^+ G^3} \approx  \frac{t}{v^2}
 -   g_h^2 \frac{t}{v^2} \frac{t}{t-m_h^2} 
\eeq 
Above the scale of the Higgs mass we obtain $\cm_{G^+G^3\rightarrow
G^+ G^3} \sim t/f_h^2$.  The amplitude continues growing
up to $\mkk$ where some other physics (e.g. vector resonances)
must restore unitarity. 
The Higgless case corresponds to $g_h = 0$.  

As recently discussed \cite{gigrpo}, this kind of behaviour is
expected on purely low energy grounds if the Higgs doublet arises as a
pseudo-Goldstone boson. Consider the case of $SO(5)$ broken to $SO(4)$
where we identify the remaining $SO(4)$ with the approximate
$SU(2)_L\times SU(2)_R$ custodial symmetry of the SM. The four
Goldstone bosons are identified with the Higgs doublet. The Lagrangian
at lowest order is:
\begin{equation}
  (1/2)\pa_\mu U^T \pa_\mu U
\end{equation}
with $U$ parameterising the Goldstone bosons:
\begin{equation}
U=f_h \exp\left(\frac{i\sqrt{2}H_a T_a}{f_h}\right) \left(
\begin{array}{c}\vec{0} \\ 1\end{array} \right)
\end{equation}
Here $T_a$ are the four broken generators of $SO(5)$ and $H_a$ are the
four Goldstone bosons of the broken symmetry that we identify with the
Higgs fields. Finally, $f_h$ is the scale of the global symmetry
breaking.
The Higgs field gets a vev $\la H_4 \ra = \ti v$. 
The electroweak breaking scale is related to its vev by  $v = f_h \sin (\ti v/f_h)$.   

To calculate scattering amplitudes we need to extract the three point
couplings of two Goldstone bosons to the Higgs boson.  Parameterising
the Higgs field as in \eref{smhnlp} and expanding the lowest order
Lagrangian we find the relevant Higgs-Goldstone couplings:
\begin{eqnarray}
\mathcal{L}_{G^2h} &=& 
\cos\left(\frac{\ti{v}}{f_h}\right) \frac{h}{v}\left(\partial_\mu G_a\right)^2
\end{eqnarray} 

Thus $g_h = \cos({\ti{v}}/{f_h})$.  The four-point couplings remain unchanged from
(\ref{eq:G^4SM}). 
Therefore, above the Higgs boson mass the WZ amplitude
grows as $\sim  (1 - \cos^2({\ti{v}}/{f_h}))t/v^2 = t/f_h^2$. 
 This is the simplest setup in which the non-unitary behaviour is encountered,
irrespectively whether the high-energy UV completion that restores
unitarity is perturbative or strongly coupled.

In the following sections we shall study 5D warped models from the
point of view of gauge boson scattering.  We will find similar
qualitative behaviour, even when the Higgs is {\it not} a
pseudo-Goldstone boson.  We will also study in detail how the
vector resonances restore unitarity of the scattering amplitudes.

\section{Holographic composite Higgs}
\label{sec:strongHolo}
\setcounter{equation}{0} 

We first consider a 5D theory with the gauge symmetry $SU(3)_c\times SU(2)_L\times SU(2)_R \times U(1)_X$ and the Higgs field on the IR brane, which was proposed in ref. \cite{agdema}. 
The limit where the brane Higgs vev goes to infinity corresponds to the Higgsless theory of ref. \cite{csgrpi}. 
The rationale behind extending the SM $U(1)_Y$ to $SU(2)_R \times U(1)_X$ is to avoid excessive contributions to the T parameter \cite{agdema,defa}.

The 5th dimension is warped with a gravitational background described
by the line element:
\begin{equation}
ds^2=a^2(x_5)\eta_{\mu\nu}dx^\mu dx^\nu-dx_5^2
\end{equation}
where $a(x_5)$ is a warp factor normalized such that $a(0)=1$.  The
5th dimension is bounded by two branes, an IR brane at $x_5=L$ and a
UV brane at $x_5=0$.  The choice $a(x_5)=1$ corresponds to flat space,
whereas $a(x_5)=e^{-k x_5}$ corresponds to AdS$_5$.  We do not specify
the warp factor in what follows other than to require that it generate
a sufficient hierarchy between the UV and the IR brane: $a(L) \equiv a_L  \ll 1$.
Moreover, we will assume that the size $L$ of the extra dimension is
large in the sense $L {\pa_5 a(L) \over a_L} \gg 1$. 
This is typically the case in backgrounds that solve the hierarchy problem such as AdS$_5$.
The significance of this last assumption will become clear in the following. 

We allow the gauge bosons to propagate in the bulk, while the Higgs
sector is confined to the IR brane. On the UV brane we explicitly
break the gauge symmetry down to the SM gauge group.


This set-up can be interpreted as an effective description of a 4D
theory with fundamental SM gauge bosons and a strongly coupled
electroweak symmetry breaking sector with a {\it global}
$SU(3)_c\times SU(2)_L\times SU(2)_R \times U(1)_X$ symmetry.  The IR
brane Higgs boson is interpreted as a composite of the strongly
interacting sector.

The 5D action for the electroweak sector is:
\begin{eqnarray}
\nonumber S&=&\int d^4x\int^L_0dx_5\sqrt{g}\left\{ -
\frac{1}{2}\text{Tr}\left(L_{MN}L_{MN}\right) -
\frac{1}{2}\text{Tr}\left(R_{MN}R_{MN}\right) -
\frac{1}{4}X_{MN}X_{MN}\right\}\\ &+&\int
d^4xdx_5\sqrt{g_4}\delta(L)\left(
\frac{1}{4}\text{Tr}\left|D_\mu\Phi\right|^2 -
V(\Phi^\dagger\Phi)\right)
\end{eqnarray}
where $L_{MN}$ and $R_{MN}$ are the $SU(2)_{L,R}$ gauge fields
respectively, $X_{MN}$ are the $U(1)_X$ gauge fields and $\Phi$ are
the scalar fields that we identify with the Higgs. The dimensionful
couplings of $SU(3)_c\times SU(2)_L\times SU(2)_R \times U(1)_X$ will
be denoted as $g_L\sqrt{L}$, $g_R\sqrt{L}$ and $g_X\sqrt{L}$.

The Higgs field acquires a vev
$\left<\Phi\right>=a_L^{-1}\tilde{v}I_{2\times 2}$ that spontaneously breaks $SU(2)_L\times
SU(2)_R\rightarrow SU(2)_V$ on the IR brane. We separate out the
physical Higgs boson from the Goldstone bosons by parameterising the
Higgs boson non-linearly:
\begin{eqnarray}
\nonumber \Phi&=&a_L^{-1}\left(\tilde{v}+h\right)U\\ \text{with
}U&=&\exp\left(\frac{i \ti G_a\sigma_a}{\tilde{v}}\right) =
\cos\left(\frac{\ti G}{\tilde{v}}\right) +
i\frac{\ti G_a\sigma_a}{\ti G}\sin\left(\frac{\ti G}{\tilde{v}}\right)
\end{eqnarray}
where
$\ti G^2=\ti G_a \ti G_a$ and $h$ is the physical Higgs boson.  From
the covariant derivative of the Higgs field we get the quadratic
terms:
\begin{equation}
\mathcal{L} = \frac{1}{2}\delta(L)\left[
\partial_\mu \ti G_a - \frac{\sqrt{L}\tilde{v}}{2}
\left(g_L L_{\mu,a}   -          g_R R_{\mu,a}\right)     \right]^2
+ \frac{1}{2}\delta(L) (\pa_\mu h)^2
 \end{equation}
These terms provide a kinetic term for the Higgs boson and the Goldstone bosons as well as a brane mass term for the gauge bosons.
They also introduce the mixing between the Goldstone bosons and the gauge bosons.

\subsection{Mass eigenstates}

The dynamics of the model can be neatly studied in the mass eigenstate
formalism introduced in \cite{aa,defa}.
 We expand the 5D fields in the mass eigenstate basis:
\begin{eqnarray}
  L_\mu^a(x,x_5) = A_{\mu,n}(x) f_{L,n}^a(x_5) &\quad& L_5^a(x,x_5) =
  G_{n}(x) \bar f_{L,n}^a(x_5)
  \nn
  R_\mu^a(x,x_5) = A_{\mu,n}(x) f_{R,n}^a(x_5)
  &\quad&
  R_5^a(x,x_5) = G_{n}(x) \bar f_{R,n}^a(x_5)
  \nn
  X_\mu(x,x_5) = A_{\mu,n}(x) f_{X,n}(x_5)
  &\quad&
  X_5(x,x_5) =  G_{n}(x) \bar f_{X,n}(x_5)
  \nn 
  && \ti G_n(x)  =  G_{n}(x) \ti f_n  
\end{eqnarray}  
where the index $n$ runs over all mass eigenstates in the theory.
The profiles $f_{n}(x_5)$ will be chosen such that the gauge bosons
are indeed mass eigenstates.  The Goldstone profiles $\bar f_{n}(x_5)$
will be chosen such that $G_n$ becomes the Goldstone boson
corresponding to the massive eigenstate $A_{\mu,n}$.  In other
words, the goal is to rewrite the quadratic part of the 5D action as a
4D action that is diagonal in $n$:
\begin{equation}
\label{e.5daq}
S_5 = \int d^4 x \sum_n \left \{ - \frac{1}{4} (\pa_{\mu}A_{\nu,n} -
\pa_{\nu} A_{\mu,n})^2 + \frac{1}{2} (\pa_\mu G_n - m_n A_{\mu,n})^2
\right \} + \text{interactions}\, .
\end{equation}
In this way, there is no tree-level mixing between the light modes and
the heavy KK modes, even in the presence of electroweak symmetry
breaking on the brane.  This is different from the more common
approach, where the KK expansion is performed in the absence of
electroweak symmetry breaking, and the electroweak symmetry breaking
leads to mixing of zero modes with KK modes.

We also retain the Goldstone degrees of freedom.  The Goldstones,
$G_n$, allow us to maintain explicit gauge invariance in the presence
of the mass term for the vector $A_{\mu,n}$.  Keeping Goldstones is
convenient as, via the equivalence theorem, scattering of
longitudinally polarized vector bosons $A_{\mu,n}$ is equivalent to
scattering of $G_n$.
   

In order to end up with the diagonal action \erefn{5daq} the profiles
$f_n(x_5)$ must satisfy:
\begin{enumerate}
\item The {\it equation of motion} 
\begin{equation}
\label{e.geom} 
\left [\pa_5 (a^2 \pa_5) +  m_n^2 \right ] f_n(x_5) = 0
\end{equation}
\item The {\it orthonormality condition} 
\begin{equation}
\label{e.gpon} 
\int_0^L  f_n(y) f_m(y) = \delta_{nm} 
\end{equation}
\item  The {\it UV boundary conditions}:
\begin{eqnarray}
\label{e.wcmbcuv} 
\pa_5 f_{L,n}^a(0) &=& 0  \qquad  a = 1,2,3 
\nn
f_{R,n}^i(0) &=& 0 \qquad  i = 1,2 
\nn
s_x \pa_5 f_{R,n}^3(0) + c_x \pa_5 f_{X,n}(0) &=& 0 \qquad  s_x = \frac{g_X}{\sqrt{g_X^2 + g_R^2}}  
\nn
-c_x f_{R,n}^3(0) + s_x f_{X,n}(0) &=& 0 \qquad  c_x = \frac{g_R}{\sqrt{g_X^2 + g_R^2}} 
\end{eqnarray}
which break $SU(2)_R \times U(1)_X$ down to $U(1)_Y$.\footnote{The
linear combination $B_\mu = s_x R_\mu^3 + c_x X_\mu$ survives on the
UV brane and its zero mode is identified with the hypercharge gauge
boson.  $B_\mu$ couples to matter with the coupling $g_Y = g_X
g_R/\sqrt{g_X^2 + g_R^2}$ and the hypercharge depends on the $SU(2)_R
\times U(1)_X$ quantum numbers via $Y = t_R^3 + X$.}
\item  The {\it IR boundary conditions} 
\begin{eqnarray}
\label{e.wcmbcir} 
\pa_5 f_{X,n}(L) &=& 0 
\nn
 g_R \pa_5 f_{L,n}^a(L) + g_L \pa_5 f_{R,n}^a(L)  & = & 0 
\nn
g_L \pa_5 f_{L,n}^a(L) - g_R \pa_5 f_{R,n}^a(L)  & = & 
-    {1 \over 4} (g_L^2 + g_R^2) a_L^{-2} L \ti v^2 (g_L f_{L,n}^a(L)  - g_R f_{R,n}^a(L)) 
\nn 
\ti f_n^a  &= &
{\sqrt L \ti v \over 2 m_n }  \left (g_L f_{n,L}^a(L) - g_R f_{n,R}^a(L) \right )
 \end{eqnarray}
\end{enumerate} 
Finally, the Goldstone profiles are determined by the gauge profiles,  
\begin{eqnarray}
\bar f_{n}(x_5) &=& m_n^{-1} \pa_5  f_{n}(x_5)\text{, for }m_n\neq 0
\nn
\bar f_{n}&=& 0\text{, for }m_n=0
\end{eqnarray}  


To calculate the explicit form of the profiles $f_n$, we solve the
equation of motion such that it satisfies the conditions (2)-(4). Instead
of solving it in a specific background it is more
convenient to proceed in a background independent fashion.  The
equation of motion is second order so it has two independent solutions that
correspond to warped trigonometric functions $C(x_5,m_n)$ and
$S(x_5,m_n)$.\footnote{The properties of warped sines $S$ and cosines
$C$ are discussed at more length in \aref{wt}.  In the flat space they
are the well-known trigonometric functions, while in \ads\, they can
be expressed as combinations of Bessel and Neumann functions, see
\eref{csads}.}  We have freedom to choose them such that they satisfy
$C(0,m_n) = 1$, $S'(0,m_n) = m_n$, $C'(0,m_n) = S(0,m_n) = 0$.  Then
the profiles can be succinctly written:
\begin{eqnarray}
f_{L,n}^a(x_5) &=& \alpha_{L,n}^a C(x_5,m_n) \nn f_{R,n}^i(x_5) &=&
\alpha_{R,n}^i S(x_5,m_n) \nn f_{R,n}^3(x_5) &=& \alpha_{N,n} s_x
C(x_5,m_n) - \alpha_{D,n} \, c_x S(x_5,m_n) \nn f_{X,n}(x_5) &=&
\alpha_{N,n} c_x C(x_5,m_n)+ \alpha_{D,n} \, s_x S(x_5,m_n) 
\end{eqnarray}

In this way, the profiles satisfy the UV boundary conditions
\erefn{wcmbcuv}.
Inserting these expressions into the IR boundary
conditions \erefn{wcmbcir} we obtain relations between the
normalization constants $\alpha_n$. 
We also obtain the quantization condition that factorizes  as $F_W(m) F_\gamma(m) F_Z(m) = 0 $. 
This gives rise to three separate classes of solutions that we refer to as the $W$, $\gamma$ and $Z$ towers (because the lightest solution will be  identified with the W, the photon and the Z, respectively). 
 The quantization conditions for these three towers read:
\begin{eqnarray}
0 &=& S'(L,m_{W,n}) C'(L,m_{W,n})
\nn &+& {a_L^{-2} L \ti v^2 \over 4} \left (
g_L^2 S'(L,m_{W,n}) C(L,m_{W,n}) + g_R^2 S(L,m_{W,n}) C'(L,m_{W,n})
\right ) \label{e.cmwtq}
\\
\label{e.cmptq}
 0&=&C'(L,m_{\gamma,n}) 
 \\ 
 0 &=& S'(L,m_{Z,n}) C'(L,m_{Z,n}) \nn 
&+& { a_L^{-2} L \ti v^2 \over 4} \left( g_L^2 S'(L,m_{Z,n}) C(L,m_{Z,n}) 
+ g_R^2 S(L,m_{Z,n}) C'(L,m_{Z,n}) + m_{Z,n} a_L^{-2} g_Y^2 \right )
\label{e.cmztq}
\nn
\end{eqnarray}

We give the exact profiles in \ref{app:boundCond1}.
In the main body of the paper we restrict ourselves to approximate expressions that are sufficient for our purposes.

\subsection{Scales}

From the quantization conditions (\ref{e.cmwtq})-(\ref{e.cmztq}) we
can extract the mass spectrum. The model contains a tower of
resonances starting at $\sim \mkk$ where\footnote{
The scale $\mkk$ gives the  {\it parametric}  dependence of the mass of the lightest vector resonances. 
In 5D Minkowski the first KK photon mass is exactly equal to $\mkk$, while in \ads\, it is approximately $0.8 \mkk$.} 
\beq
\label{e.kks}
\mkk = {\pi \over \int_0^L a^{-1}(y)} \sim \pi a'(L) 
\eeq

There are two exceptions. Firstly, as $U(1)_{em}$ is unbroken, there is
always a massless vector boson - the photon.  Secondly, there can be
further states with masses parametrically below $\mkk$ that are
identified with the W and Z bosons.
In \eref{cssz} we calculate their masses by expanding the warped trigonometric functions in
\eref{cmwtq}, (\ref{e.cmztq}) at small $m$.

From these general results, we now consider two explicit limiting
cases. 
First consider the limit where $\ti v \gg \mkk$. In this case the quantization conditions \erefn{cmwtq} and \erefn{cmztq} are dominated by the second term. In such a limit we obtain
\begin{equation}
m_W^2  \approx  {g_L^2 f_h^2 \over 4} 
\qquad 
m_Z^2 \approx {(g_L^2 + g_Y^2 ) f_h^2 \over 4}  
\end{equation}
where the scale $f_h$, called  the Higgs decay constant,  is fixed by the geometry of the 5th dimension: 
\begin{equation} 
\label{e.cmf}
f_h^2 =  {4 \over L (g_L^2 + g_R^2) \int_0^L a^{-2}(y)}
\end{equation}
Thus in this limit $v \approx f_h$ and, as we will see later,  $g_h=0$. 
Therefore the Higgs plays no role in unitarising the gauge boson scattering, 
though it remains in the spectrum.\footnote{
Generically, in this limit we expect the Higgs boson to be heavy, $m_h \sim \ti v$, but if its self-coupling is very weak it may remain light.}
We refer to this limit as the {\it Higgsless limit}.

If we take $g_L \sim g_R$, we get:
\begin{equation}
\label{e.cmvf}
g_L^2 f_h^2 \sim {4 \mkk^2 \over \pi^2 \cv} \qquad \qquad 
\cv \equiv  {L  a'(L) \over a_L} 
\end{equation}
  
We have introduced the geometric factor $\cv$ that we call the {\it
volume factor}. In the Higgsless limit we need the volume factor to be
large, otherwise there is no separation between $v$ and $\mkk$. Such a
case is ruled out by searches for light resonances.  In flat space we
get $\cv = 1$. In contrast, within the Randall-Sundrum \ads\ setup it
is $\cv = k L \sim \log(\mpl/\mkk) \sim 30$.  As discussed in
\cite{defa}, we expect $\cv$ to be large if the 5D set-up is to solve
the hierarchy problem.

The volume factor could be made arbitrarily large by an educated choice of the 5D geometry.   
Note however that consistency arguments set an upper bound, $\cv \simlt 16 \pi$. 
Otherwise, the resonance scale would be pushed above 1 TeV and the gauge boson scattering would get strong before the vector resonances set in to restore unitarity.  

Now consider the opposite limit in which $\tilde{v}\ll
\mkk$.\footnote{The condition $\tilde{v}\ll \mkk$ is a postulate, but
in general this input is a consequence of some unspecified dynamics
that gives rises to the boundary Higgs potential.  We take $\ti
v/\mkk$ as a free parameter and set it to be small but getting to
generate $\ti v/\mkk < 1/4\pi$  would typically require fine-tuning.} In this
case $g_h\approx 1$ and the Higgs boson is SM-like. We refer to this
limit as the {\it Higgs limit}. Once again, this allows us to obtain
the desired scale separation $v \ll\mkk$. 
For $\ti v/\mkk \ll 1$ the
electroweak gauge boson masses are approximately given by
\begin{equation}
m_W^2 \approx {g_L^2 \ti v^2 \over 4} 
\qquad \qquad 
m_Z^2 \approx {g_L^2 \ti v^2 \over 4 \cos^2 \theta_W}  
\end{equation} 
Thus, the electroweak scale in the Higgs limit is simply $v \approx
\ti v$.  More precisely, the relation between the two scales is of the
form $v^2 \approx \ti v^2 \left ( 1 - {\ti v^2 \over f_h^2} \right)$.
The scale $f_h$ that appears in this relation is the one defined by eq. (\ref{e.cmf}). 
Although in the Higgs limit $f_h$ enters in a rather intricate way, it will turn out to play an important role in describing the $W$ and $Z$ scattering.

Summarizing, the following three scales have emerged: the electroweak
scale $v$, the Higgs decay constant $f_h$ and the resonance scale
$\mkk$.  The separation between $f_h$ and $\mkk$ is set by a geometric
quantity we call the volume factor.  In the Higgs limit the
electroweak scale can be adjusted to be smaller than $f_h$, while in
the Higgsless limit $v$ and $f_h$ coincide.

\subsection{Goldstone bosons}

The important input for calculating gauge boson scattering amplitudes
are the profiles of the Goldstone bosons $G^+$, $G^-$ and $G^3$ that
are eaten by the physical $W^+$, $W^-$ and $Z$ bosons respectively.
In general, they are linear combinations of $L_5$, $R_5$, $X_5$ and
$\ti G$. 
The exact profiles are given in \eref{cmegp}. 
To leading order in $m_W/\mkk$, these profiles can be
concisely described in a background independent way. 
With the help of \eref{cssz} we find
\begin{eqnarray}
\label{e.cmwsis} 
\bar f_{L,W}^i &\approx&  - {1 \over \sqrt L} m_W x_5 a^{-2}(x_5)
\nn
\bar f_{R,W}^i &\approx&  {1 \over \sqrt L} {g_R \over g_L} m_W L a^{-2}(x_5) 
\nn
\ti f_W^i &\approx& {v \over \ti v}
\end{eqnarray} 
\begin{eqnarray}
\label{e.cmzsis} 
\bar f_{L,Z}^3 &\approx& - {1 \over \sqrt L} m_W x_5 a^{-2}(x_5) 
\nn
\bar f_{R,Z}^3 &\approx& {1 \over \sqrt L}  {g_R \over g_L} \left ( s_x^2 x_5 + c_x^2 L \right ) m_W a^{-2}(x_5) 
\nn
\bar f_{X,Z} &\approx& c_x {\tan \theta_W \over \sqrt L}  \left ( x_5 -  L \right ) m_W a^{-2}(x_5)  
\nn
\ti f_Z^3 &\approx& {v \over \ti v}   
\end{eqnarray} 
In the Higgs limit, $v \approx \ti v$ and the Goldstones reside mostly on the brane. 
In the Higgsless limit, $\ti v \gg v$ and the Goldstones live mostly in the bulk, though for warped
metrics they are still sharply localized at the IR brane due to the $a^{-2}(x_5)$ profile.

With this information, we can read off the quartic self-couplings of
the Goldstones and the Higgs-Goldstone couplings.  We find 
\begin{eqnarray}
\nn  \cl_{G^4}&=& - {g_h^2 \over 6 v^2} (\pa_\mu G^a)^2 G^2 +  {g_h^2 \over 6 v^2} (\pa_\mu G^a G^a)^2 \\
  \cl_{G^2 h} &=& {g_h \over v} h  (\pa_\mu G^a)^2  \label{e.gmi}
\end{eqnarray} 
where $g_h = v^3/\ti v^3$.
Thus, $g_h \approx 0$ in the Higgsless limit, while $g_h \approx 1$ in the Higgs limit, as indicated before.

\subsection{Couplings of resonances}
\label{s.cmcr}

To calculate the total gauge boson scattering amplitude we need to
calculate the coupling of the Goldstone bosons to the resonances.  The
Goldstones interact with the charged and neutral resonances via the
triple vertices:
\begin{eqnarray}  
\cl_{G^2 A} &=& - i (\pa_\mu G^+ G^- - \pa_\mu G^- G^+) g_{N,n} A_{\mu,n}
\nn
 & +& \left \{ - i (\pa_\mu G^- G^3 - \pa_\mu G^3 G^-)
g_{C,n} W_{\mu,n}^+ + \hc \right \}
\label{e.grc} 
\end{eqnarray}
where $N$ stands for neutral bosons $\gamma,Z$ and $C$ stands for the
charged $W$.  The resonance couplings $g_n$ can be found by inserting
the Goldstone and resonance profiles into the interaction terms in the
5D action.  In order to somewhat simplify the resulting expressions we
set $g_L = g_R$.  Furthermore, the Goldstone profiles are localized
toward the IR brane. This results in the factor $a^{-2}(y)$ showing up in
the integrals. Therefore the integrals are dominated by the IR region
and it is a good approximation to replace $y \approx L$ in the
integrands.  This allows us to approximate the resonance couplings as
\begin{eqnarray} 
g_{N,n} &\approx& g_L m_W^2 L^{3/2} \int_0^L a^{-2}(y) \left \{
 f_{L,n}^3(y) + f_{R,n}^3(y) \right \} \nn & +&  
  {1 \over 2} \sqrt{L}
 g_L {v^2 \over \ti v^2} (f_{L,n}^3(L) + f_{R,n}^3(L)) \\ \nn g_{W,n}
 &\approx& g_L {m_W^2} L^{3/2} \int_0^L a^{-2}(y) \left \{
 f_{L,n}^i(y) + f_{R,n}^i(y) \right \}
 \nn
  &+& {1 \over 2} \sqrt{L} g_L
 {v^2 \over \ti v^2} (f_{L,n}^i(L) + f_{R,n}^i(L))
\label{e.cmrca} 
\end{eqnarray}
The first term within the integral dominates the Higgsless limit, $v
\ll \ti v$, when the Goldstones live in the bulk.  The second term
dominates the Higgs limit, when the Goldstones live on the brane.  For
$n$ corresponding to the electroweak gauge bosons we recover the
Standard Model couplings $g_0$: $g_W \approx {g_L \over 2}$, $g_Z
\approx {g_L^2 - g_Y^2 \over 2 \sqrt{g_L^2 + g_Y^2}}$, $g_\gamma = e$.
For $n$ corresponding the heavy resonances the results are collected
in \eref{cmrcwzg}.

In order to estimate the resonance couplings we can employ the approximation (\ref{e.wtwr}) to the profiles of resonances. 
The simple pattern that emerges is that all resonance
couplings, $g_n$, are parametrically enhanced with respect to the SM ones by a common
factor:
\begin{equation}
\label{e.cmpc}
g_n \sim \sqrt{\cv} g_0
\end{equation}  
where $\cv$ is the volume factor defined in \eref{cmvf} and $g_0$ is
the SM coupling relevant for the given tower: $g_W$, $g_Z$ or
$g_\gamma$.  Since $\cv \sim \mkk^2/f_h^2$, we get $f_h g_n \sim \mkk
g_0$.  For $g_0 \sim 1$ this coincides with the relation advertised in
eq. (1) of ref. \cite{gigrpo}.

To make the discussion more quantitative, consider first the couplings of the charged resonances in the Higgs limit.
In this case the quantization condition for the $W$ tower \erefn{cmwtq} reduces to
$C'(L,m_{W,n}) S'(L,m_{W,n}) \approx 0$.
Thus, the mass eigenstates in the W tower split into Neumann-Neumann (NN) and Dirichlet-Neumann
(DN) modes (\ref{Wtower1}):
\begin{eqnarray}
  ({\rm NN}): \quad f_{L,n}^i &\approx& {C(x_5,m_{W,n}) \over \left
    (\int_0^L [C(y,m_{W,n})]^2 \right )^{1/2}} \qquad C'(L,m_{W,n}) =
    0\\ \nn ({\rm DN}): \quad f_{R,n}^i &\approx& {S(x_5,m_{W,n})
    \over \left (\int_0^L [S(y,m_{W,n})]^2 \right )^{1/2}} \qquad
    S'(L,m_{W,n}) = 0
\end{eqnarray}
 For warped metrics, the UV boundary conditions are not relevant for
the behaviour of the resonance profiles in IR (they only affect the
light modes that are delocalized).  More precisely, we have the
relations 
\beq
C(L,m) \approx - m L S(L,m) \qquad   C'(L,m) \approx - m L S'(L,m)
\eeq
that hold for $m \sim \mkk$. 
In consequence, the NN and the DN modes have approximately the same
masses and, from \eref{cmrca}, couple with approximately the same
strength to the electroweak Goldstone bosons.  Thus, at the scale of
the first resonance the Goldstone bosons couple to two almost
degenerate charged vector states.  Moreover, using the methods of
\cite{hisa_sr}, one can prove some remarkable sum rules.  A profile
satisfying the DN boundary conditions has the integral representation
$f_n(y) = m_n^2 \int_0^y a^{-2}\int_{y'}^L f_n(y'')$.  Thus 
\beq
\sum_n {f_n(L)^2 \over m_n^2} =\sum_n f_n(L) \int_0^L
a^{-2}\int_{y'}^L f_n(y'') = \int_0^L a^{-2} 
\eeq where we have used
the completeness relation $\sum_n f_n(x) f_n(y) = \delta(x-y)$.  For
NN modes we get approximately the same sum rule, once the
contribution of the SM $W$ boson is omitted in the sum.  Using these
results we easily obtain \beq
\label{e.cmsrh}
\sum_{n>0} {g_{W,n}^2 \over m_{W,n}^2}  \approx {1 \over f_h^2}
\eeq 
This  sum rule is typically dominated by the first two terms, so we find
\beq
g_{W,1}^2 \approx  g_{W,2}^2 \approx {m_{W,1}^2 \over 2 f_h^2}
\eeq  
This is in accord with the parametric estimate \erefn{cmpc}.  

The Higgsless limit, although qualitatively similar, differs in several details. 
The $W$ tower splits now into the vector and axial modes:   
\bea
f_{L,n}^i &\approx& {C(x_5,m_{W,n})  \over  \left (2 \int_0^L   [C(y,m_{W,n})]^2 \right )^{1/2}}
\quad f_{R,n}^i \approx f_{L,n}^i \qquad C'(L,m_{W,n}) =   0 
\nn 
f_{L,n}^i &\approx& {C(x_5,m_{W,n})  \over  \left (2 \int_0^L   [C(y,m_{W,n})]^2 \right )^{1/2}}
\quad f_{R,n}^i \approx - f_{L,n}^i 
\qquad S(L,m_{W,n}) = 0
\eea 
From \eref{cmrca}, only the vector modes couple to the electroweak Goldstones.
The sum rule now becomes 
\beq
\label{e.cmsrhl}
\sum_{n>0} {g_{W,n}^2 \over m_{W,n}^2}  \approx {1 \over 3 f_h^2}
\eeq    
At the scale of the first resonance only one vector state appears in the Goldstone scattering amplitude. 
Its coupling can be estimated as 
\beq
g_{W,1}^2 \approx {m_{W,1}^2 \over 3 f_h^2} ,
\eeq 
so that it is slightly weaker than in the Higgs limit. 

To complete the picture let us discuss the cutoff scale where the 5D
theory becomes strongly coupled.  From the parametric dependence of
the resonance couplings we conclude the cutoff is fixed by the volume
factor.  We can estimate:
\begin{equation}
\Lambda \sim {4 \pi \over g_0^2 \cv} \mkk   
\end{equation}
We see the volume factor should not be larger than $\sim 16 \pi$.
Otherwise, the first resonance would already be strongly coupled and
the 5D description would not be meaningful.

\section{Holographic pseudo-Goldstone Higgs}
\label{s.hpgh}
\setcounter{equation}{0} 

We move to another, closely related higher-dimensional setup.  We
consider a 5D $SU(3)_C \times SO(5)\times U(1)_X$ gauge theory broken
to $SU(3)_C \times SU(2)_L\times U(1)_Y$ on the UV brane, and to
$SU(3)_C\times SO(4)\times U(1)_X$ on the IR brane \cite{agcopo}. The
larger symmetry group of the bulk allows us to accommodate the Higgs
field as the 5th component of the gauge bosons.  The Higgs field is
massless at tree level due to 5D gauge invariance, but it acquires a
potential at one loop.  Thus, the origin of the light Higgs field is
addressed in this set-up (rather than postulated, as in the previous
one).  In a fully fledged theory the Higgs field would acquire a vev
dynamically through minimization of the Coleman-Weinberg potential.
Here however, we will not study the dynamics that produces the vev,
but simply assume it exists.  Holographically, this setup again
corresponds to the Standard Model coupled to a strongly interacting
sector that breaks electroweak symmetry.  The global symmetry of the
strong sector is $SU(3)_C \times SO(5)\times U(1)_X$.  $SO(5)$ is
spontaneously broken to $SO(4)$ by the strong dynamics.  The resulting
pseudo-Goldstones are identified with the Higgs field.

We concentrate on the $SO(5)\times U(1)_X$ part with the gauge fields
$A_M = A_M^\alpha T^\alpha$ and $X_M$.  The dimensionful bulk gauge
couplings are denoted as $g \sqrt{L}$ and $g_X \sqrt{L}$.  The 5D
action is
\begin{equation}
S_{\mathrm 5 \mathrm D}= \int d^4 x \int_0^L dx_5 \sqrt{g} \left (
- {1 \over 4} \tr \{ A_{MN} A^{MN} \}    -    {1 \over 4}  X_{MN} X^{MN} 
\right )  \, ,  
\end{equation}

\subsection{Mass eigenstates}

We employ the mass eigenstate formalism for the KK expansion.  We want
to arrive at the quadratic action of the form (\ref{e.5daq}) which is
diagonal in the KK index in the presence of electroweak symmetry
breaking. In contrast to the previous section there is an added
complication of the $A_5$ vev which affects the quadratic terms in
the action.  The changes can, however, be simply taken into account by
replacing $\pa_5$ with the covariant derivative 
$D_5 = \pa_5 - i g \sqrt{L} [\la A_5 \ra, \cdot]$.

We perform the KK decomposition, in the presence of the $A_5$ vev: 
\begin{eqnarray}
A_\mu(x,x_5) &=& A_{\mu,n}(x) f_{n}(x_5, \la A_5 \ra )
\quad
A_5(x,x_5) = \la A_5(x_5) \ra + G_{n}(x) \bar f_{n}(x_5, \la A_5\ra )
\nn
X_\mu(x,x_5) &=& A_{\mu,n}(x) f_{X,n}(x_5, \la A_5 \ra )
\quad
X_5(x,x_5) =  G_{n}(x) \bar f_{X,n}(x_5, \la A_5 \ra )
\end{eqnarray}  
where $f_n = f_n^\alpha T^\alpha$.  We split the $SO(5)$ generators as
$T^\alpha = (T_L^a, T_R^a, T_C^\ha)$, $a = 1\dots 3$, $\ha = 1 \dots
4$, corresponding to $SU(2)_L$ and $SU(2)_R$ subgroups and the
$SO(5)/SO(4)$ coset.  Accordingly, we also split the gauge field $A_M
= (L_M,R_M,C_M)$ and the profiles $f_{n} = (f_{L,n},f_{R,n},f_{C,n})$.

Diagonalization is achieved when the profiles satisfy the following conditions:

\begin{enumerate}
\item The equation of motion in the $A_5$ background:
  \begin{eqnarray}
    \nonumber    &   D_5(a^2 D_5 f_n) + m_n^2 f_n = 0&\\ 
    &   D_5 f_n = \pa_5 f_n - i g \sqrt{L} [\la A_5(x_5) \ra,f_n]  &
  \end{eqnarray} 

$f_n^X$ satisfies the same equation with $D_5 \to \pa_5$.
\item The normalization condition:
\begin{equation}
\int_0^L \left \{ 
\tr[f_n(y, \la A_5 \ra) f_n(y, \la A_5 \ra)] + f_n^X(y, \la A_5 \ra) f_n^X(y, \la A_5 \ra)
\right \} = 1
\end{equation} 
\item IR boundary conditions:
  \begin{eqnarray}
    \nonumber f_{C,n}^\ha(L, \la A_5 \ra ) &=& 0\\ \nonumber D_5 f_{L,n}^a(L, \la A_5 \ra )
    &=& 0\\ \nonumber D_5 f_{R,n}^a(L, \la A_5 \ra ) &=& 0\\  \pa_5 f_{X,n}(L,
    \la A_5 \ra ) &=& 0
  \end{eqnarray} 
  that break $SO(5)\times U(1)_X$ to $SO(4)\times U(1)_X$.  
\item UV boundary conditions:
\begin{eqnarray}
\label{e.hpguvb}
\pa_5 f_{L,n}^a(0, \la A_5 \ra ) &=& 0  
\nn
f_{R,n}^i(0, \la A_5 \ra ) &=& 0 \qquad  i = 1,2 
\nn
s_x \pa_5 f_{R,n}^3(0, \la A_5 \ra ) + c_x \pa_5 f_{X,n}(0, \la A_5 \ra ) &=& 0 \qquad  s_x = {g_X \over \sqrt{g_X^2 + g_L^2}}  
\nn
-c_x f_{R,n}^3(0, \la A_5 \ra ) + s_x f_{X,n}(0, \la A_5 \ra ) &=& 0 \qquad  c_x = {g_L \over \sqrt{g_X^2 + g_L^2}} 
\nn
f_{C,n}^\ha(0, \la A_5 \ra ) &=& 0 \qquad  \ha = 1..4 
\end{eqnarray}
that break $SO(5)\times U(1)_X$ to $SU(2)_L\times U(1)_Y$, the
hypercharge being a linear combination of $SU(2)_R\times U(1)_X$.  The
SM gauge couplings are $ g_L = g$ and $g_Y = {g_X g \over \sqrt{g_X^2
+ g^2}}$
\end{enumerate}

The Goldstone profiles are chosen accordingly :
\begin{equation}
  \bar f_n(x_5, \la A_5 \ra ) = m_n^{-1} D_5 f_n(x_5, \la A_5 \ra) 
\end{equation}  

In general, the profiles in the $A_5$ background are related to the
profiles at $\la A_5 \ra = 0$ by a rotation via the Wilson-line
matrix,
\begin{eqnarray}
  \label{e.vrt} 
    f_n(x_5, \la A_5 \ra) &=& \omega^{-1}(x_5, \la A_5 \ra) f_n(x_5)
  \omega(x_5, \la A_5 \ra),\\ \nn \text{where }&& \omega = P \exp
  \left (- i g \sqrt{L} T^\alpha \int_0^{x_5} \la A_5^\alpha \ra
  \right )
\end{eqnarray}
and $f_n^X(x_5, \la A_5 \ra) = f_n^X(x_5)$.  The profiles $f_n(x_5)$
satisfy the ``normal'' equation of motion, $\pa_5(a^2 \pa_5 f_n) + m_n^2
f_n = 0$.  Similarly, for the Goldstone profiles
\begin{equation}
\label{e.gfh} 
\bar f_n(x_5, \la A_5 \ra) = m_n^{-1} \omega^{-1}(x_5, \la A_5 \ra)
\pa_5 f_n(x_5) \omega(x_5, \la A_5 \ra)
\end{equation}

In the following we choose the basis such that the vev resides in only
one direction in the group space:
\begin{equation}
\la A_5^{\hat 4} \ra = { a^{-2}(x_5) \over \sqrt {\int_0^L a^{-2}(y)}}
\ti v
\end{equation} 
The profiles at zero vev can be written as
\begin{eqnarray}
f_{L,n}^a(x_5) &=& \alpha_L^a C(x_5,m_n)
\nn
f_{R,n}^i(x_5) &=& \alpha_R^i S(x_5,m_n) \qquad i = 1,2
\nn
f_{R,n}^3(x_5) &=& \alpha_N s_x  C(x_5,m_n) - \alpha_D \,  c_x S(x_5,m_n)
\nn
f_{X,n}(x_5) &=& \alpha_N c_x  C(x_5,m_n)+ \alpha_D \, s_x S(x_5,m_n)
\nn
f_{C,n}^\ha(x_5) &=& \alpha_C^\ha S(x_5,m_n)
\end{eqnarray}
They satisfy the UV boundary conditions \erefn{hpguvb}.  The constants
are obtained by imposing the IR boundary conditions. The solutions can
be organized into two towers $W_n$, $\bar W_n$ of charged gauge
bosons, and four towers $\gamma_n, Z_n,\bar Z_n, H_n$ of neutral ones.
We list all the profiles in \ref{app:boundCond}.  Here we content
ourselves with the quantization conditions: \beq
\label{e.pgwtqc}
C'(L,m_{W,n}) S(L,m_{W,n}) 
+ {1 \over 2} m_{W,n} a_L^{-2} \sin^2\left ( \ti v \over f_h\right) = 0
\qquad
S'(L,m_{\bar W,n}) = 0  
\eeq 
\beq
\label{e.pgztqc}
\cos^2 \theta_W C'(L,m_{Z,n}) S(L,m_{Z,n}) + {1 \over 2} m_{Z,n}
a_L^{-2} \sin^2\left(\ti v \over f_h\right) = 0
\qquad 
S'(L,m_{\bar Z,n}) = 0  
\eeq
\beq
C'(L,m_{\gamma,n}) = 0   \qquad 
S(L,m_{H,n}) = 0 
\eeq  
Only the masses in the $W$ and $Z$ towers are sensitive to electroweak
symmetry breaking.  The scale $f_h$ is once again defined in
terms of geometric quantities:
\beq
f_h^2 =   {2 \over g_L^2 {L \int_0^L a^{-2}(y)}} 
\eeq 

This definition coincides with \eref{cmf}, once $g_L = g_R$.  
In the present model $f_h$ appears as the symmetry breaking scale at
which the global $SO(5)$ is broken to $SO(4)$. 
Its role in the $W$ and $Z$ scattering will turn out analogous as in the previous model. 

\subsection{Scales}

The tower of heavy resonances begins at $\sim \mkk$ (defined in
\eref{kks}). In addition the setup can accommodate light vector states
identified with the electroweak gauge bosons. There is always the
photon with $m_\gamma = 0$.  The lightest massive vector states in the
$W$ and $Z$ tower are identified with the $W$ and $Z$ bosons and can
be parametrically lighter than $\mkk$. Finally, there are no light
states in the remaining towers.  Expanding the warped trigs
in \eref{pgwtqc} and \eref{pgztqc} for small masses we find the $W$ and $Z$
masses:
\begin{eqnarray}
\nn m_W^2 &\approx& {g_L^2  \over 4} f_h^2\sin^2\left(\ti v \over f_h\right)
\\ 
m_Z^2 &\approx& {g_L^2 + g_Y^2 \over 4} f_h^2  \sin^2\left(\ti v \over f_h\right) 
\end{eqnarray} 
Thus we identify the electroweak scale as $v \approx f_h \sin (\ti
v/f_h)$.  The separation between the electroweak scale and the
resonance scale can be achieved in two separate limits. In the first
case we assume a separation between the Higgs vev and the decay
constant, $\sin (\ti v/f_h) \ll 1$.  This is the {\it Higgs limit}.
$\sin (\ti v/f_h)$ is a free parameter until we specify the dynamics
that gives rises to the Higgs potential.  One should be aware,
however, that getting $\ti v/f_h$ smaller than 1 typically requires
fine-tuning. In the other limit, $\sin (\ti v/f_h) \sim 1$. This
corresponds to the {\it Higgsless limit}. Once again we need to
separate $f_h$ from $\mkk$.  The separation $f_h \ll \mkk$ is obtained
in the 5D background with a large volume factor $\cv$.

Summarizing, the three scales $v$, $f_h$, and $\mkk$ have emerged
again.  In the Higgsless limit $v\approx f_h$.  In fact, in this limit
the quantization conditions for the $W$ and $Z$ tower masses are
exactly the same as in the previous model.  The physics in the
Higgsless limit is indistinguishable in these two models.

\subsection{Goldstone bosons}
To calculate the gauge boson scattering amplitudes we need to
calculate the Goldstone boson profiles corresponding to $W$, $Z$.  The
exact profiles are written in \eref{pggpe}.  Expanding these profiles
in powers of $m_W^2$ and $m_Z^2$ we find, at lowest order:
\begin{eqnarray}
\label{e.hpgwsis} 
\bar f_{L,W}^i &\approx&  - {1 \over \sqrt L} m_W x_5 a^{-2}(x_5)
\nn
\bar f_{R,W}^i &\approx&  {1 \over \sqrt L}  m_W L a^{-2}(x_5) 
\nn
\bar f_{C,W}^i &\approx&  {1 \over \sqrt L}  {\rt \cos(\ti v /f_h) \over \sin(\ti v /f_h)} m_W L a^{-2}(x_5) 
\end{eqnarray} 
\begin{eqnarray}
\label{e.hpgzsis} 
\bar f_{L,Z}^3 &\approx& - {1 \over \sqrt L} m_W x_5 a^{-2}(x_5) 
\nn
\bar f_{R,Z}^3 &\approx& {1 \over \sqrt L}  \left ( s_x^2 x_5 + c_x^2  L \right ) m_W a^{-2}(x_5) 
\nn
\bar f_{X,Z} &\approx& c_x {\tan \theta_W \over \sqrt L}  \left (x_5  -  L \right ) m_W a^{-2}(x_5)  
\nn
\bar f_{C,Z}^3 &\approx& {1 \over \sqrt L}  {\rt \cos(\ti v /f_h) \over \sin(\ti v /f_h)} L m_W a^{-2}(x_5) 
\end{eqnarray} 

Compared to the previous model, there are no boundary
Goldstones. Instead their role is taken over by $C_5^a$.  The
parameter controlling the distribution of the Goldstones is now
$\sin(\ti v /f_h)$.  In the Higgs limit the electroweak Goldstone
bosons are mainly composed of $C_5^a$.  In the Higgsless limit the
electroweak Goldstones flow to $L_5^a$ and $R_5^a$.  In all cases the
Goldstones are sharply localized on the IR brane with the profile
behaving as $a^{-2}$.
The self-interactions of the non-linearly defined Goldstones and the triple vertex with the Higgs boson are described by \eref{gmi} with $g_h = \cos(\ti v /f_h)$.\footnote{%
Suppression of the WWh and ZZh vertices by  $\cos(\ti v /f_h)$ was also pointed out in ref. \cite{hosa}.}  
In the Higgsless limit the physical Higgs decouples from the Goldstones but it remains in the physical
spectrum.

\subsection{Couplings to resonances}
\label{s.pgcr}

The couplings of the electroweak Goldstone bosons to the resonances
are given in \ref{app:boundCond}.  In the following discussion we
again make the assumption that the warp factor is steep enough close
to the IR brane, so that we can replace $y \to L$ in all integrals.
In such a case, the coupling of the vertices defined in \eref{grc} can
be written as
\begin{eqnarray}
\label{e.pgrca} 
  g_{C,n} &\approx& g_L L^{3/2} m_W^2 
  \int_0^L a^{-2} \left \{ f_{L,n}^i + f_{R,n}^i \right\}\\
 g_{N,n} &\approx & g_L  L^{3/2} m_W^2
   \int_0^L a^{-2} \left \{ f_{L,n}^3 + f_{R,n}^3 \right\}
\end{eqnarray} 
where $C = W,\bar W$ stands for charged, while $N = Z, \bar Z, \gamma$
stands for neutral (the vector bosons from the $H$ tower do not
couple to the electroweak Goldstones).  The Standard Model gauge
bosons couple to the electroweak gauge bosons as $g_W \approx g_L/2$,
$g_Z \approx {g_L^2 - g_Y^2 \over 2 \sqrt{g_L^2 + g_Y^2}}$, $g_\gamma
= e$.  The resonance couplings depend on their profiles which we
collected in \ref{app:boundCond}.  Parametrically, we again observe an
enhancement of the resonance couplings
\begin{equation}
\label{e.pgpc} 
g_n \sim \sqrt{\cv} g_0
\end{equation}  
where $g_0$ is the coupling of the corresponding Standard Model gauge
boson.  Moreover, the couplings of the $\bar W$ and $\bar Z$ towers
are also proportional to $\cos(\ti v/f_h)$, so that they decouple from
the electroweak Goldstones in the Higgsless limit.

Consider first the couplings of the charged resonances in the Higgs limit. 
In this case the quantization condition in the $W$ tower \erefn{pgwtqc} reduces to  
$C'(L,m_{W,n}) S(L,m_{W,n}) \approx 0$. 
Thus, the mass eigenstates in the $W$ tower split into Neumann-Neumann (NN) and Dirichlet-Dirichlet (DD) modes:  
\bea
({\rm NN}): f_{L,n}^i &\approx& {C(x_5,m_{W,n})  \over  \left (\int_0^L   [C(y,m_{W,n})]^2 \right )^{1/2}}
\qquad C'(L,m_{W,n})  = 0 
\nn 
({\rm DD}): f_{C,n}^i &\approx& {S(x_5,m_{W,n})  \over  \left (\int_0^L   [S(y,m_{W,n})]^2 \right )^{1/2}}
\qquad S(L,m_{W,n}) = 0
\eea 
From \eref{pgrca}, only the NN modes couple to the electroweak Goldstone bosons. 
The $\bar W$ tower has the profile of the DN type 
\bea
({\rm DN}): f_{R,n}^i &\approx& {S(x_5,m_{\bar W,n})  \over  \left (\int_0^L   [S(y,m_{\bar W,n})]^2 \right )^{1/2}}
\qquad S'(L,m_{\bar W,n})  = 0
\eea 
As discussed before the NN and the DN resonances have approximately
the same profiles and masses, therefore they couple with approximately
the same strength to the electroweak Goldstone bosons.  Thus, at the
scale of the first resonance, there are two degenerate charged vector
states.  Moreover we obtain the following sum rules:
\beq
\sum_{n} {g_{W,n}^2 \over m_{W,n}^2} 
\approx 
\sum_{n} {g_{\bar W,n}^2 \over m_{\bar W,n}^2} \approx {1 \over 6 f_h^2}
\eeq 
The sum rules are typically dominated by the first term, so we get 
\beq
g_{W,1}^2 \approx g_{\bar W,1}^2 \approx {m_{W,1}^2 \over 6 f_h^2}
\eeq  

With respect to gauge boson scattering, the Higgsless limit is indistinguishable from the previous model. 
The $\bar W$ tower decouples from the electroweak Goldstone bosons. 
The $W$ tower splits now into the vector and axial modes:   
\bea
f_{L,n}^i &\approx& {C(x_5,m_{W,n})  \over  \left (2 \int_0^L   [C(y,m_{W,n})]^2 \right )^{1/2}}
\quad f_{R,n}^i \approx f_{L,n}^i \qquad C'(L,m_{W,n}) =   0 
\nn 
f_{L,n}^i &\approx& {C(x_5,m_{W,n})  \over  \left (2 \int_0^L   [C(y,m_{W,n})]^2 \right )^{1/2}}
\quad f_{R,n}^i \approx - f_{L,n}^i 
\qquad S(L,m_{W,n}) = 0
\eea 
From \eref{pgrca}, only the vector modes couple to the electroweak Goldstones.
The sum rule now becomes 
\beq
\sum_{n} {g_{W,n}^2 \over m_{W,n}^2}  \approx {1 \over 3 f_h^2}
\eeq   
Thus, at the scale of the first resonance effectively only one vector
state appears in the Goldstone boson scattering amplitude. On the
other hand, its coupling is stronger than in the Higgs limit by
the factor $\sim \sqrt 2$:
\beq
\label{e.rce4}
g_{W,1}^2 \approx {m_{W,1}^2 \over 3 f_h^2}
\eeq 

\section{Gauge boson scattering amplitudes}
\label{s.d}
\setcounter{equation}{0} 

We come back to discussing the scattering of longitudinally polarized electroweak gauge bosons.
In the following we discuss the specific case $W_L Z_L \to W_L
Z_L$. The other scattering processes follow precisely the same logic.
Quite generally, in the 5D models the scattering amplitude of the
corresponding Goldstone fields has the form
 \begin{equation}
 \label{e.pgwzwzs} 
 \cm_{G^+ G^3 \to G^+ G^3} =  -  {g_h^2 \over v^2} {t m_h^2 \over t - m_h^2 } 
 -  \sum_n g_{C,n}^2  \left ({t - s \over u - m_{n}^2} + {t - u \over s - m_{n}^2} \right ) 
 \end{equation}
The sum runs over all charged vector boson states
and $g_{C,n}$ denotes the couplings of the electroweak Goldstone
bosons with the charged vector resonances.
The first term comes from the self-interactions of the Goldstones and
triple vertices with the physical Higgs boson, as in \eref{gmi}.  In
the 5D models we consider, the quartic coupling is always correlated
with that of the Higgs-Goldstone coupling and is given by $g_h^2/v^2$.

From the Goldstone amplitude we can extract the dominant term in the
scattering amplitude $W_L Z_L \to W_L Z_L$.
 At energies above the $m_W$ mass, but below the Higgs mass and the resonances scale, the
amplitude contains the term that grows quadratically with energy:
\begin{equation}
\label{e.pgwzwzs1} 
 \cm_{G^+ G^3 \to G^+ G^3} \approx  
\left ( {g_h^2 \over v^2}   + 3  \sum_{n>0} {g_{C,n}^2 \over m_n^2} \right ) t  \qquad t < m_h 
\end{equation}
The second term is the contribution of the heavy charged resonances
(the light $W$ is omitted in this sum) that, at low energies, induce
an effective four-Goldstone vertex.  In fact, at this order the behaviour of the
amplitude below the Higgs mass follows from the low energy theorems \cite{chgoge},  
$\cm \approx t/\rho v^2$. 
In our 5D models, $\rho \approx 1 + \co(v^2/\mkk^2)$ due to the custodial symmetry. 
 We thus conclude that the amplitude must grow like $t/v^2$.  This tells us that the contribution of the resonances should adjust appropriately and there should be the sum rule 
\beq 
3 \sum_{n>0} {g_{C,n}^2 \over m_n^2}  \approx {1- g_h^2 \over v^2} 
\eeq 
There should also be an analogous sum rule involving neutral
resonances.  In several limiting cases, we have derived these precise
sum rules through analytical calculations in 5D.

At energies above the Higgs mass, the first term in \eref{pgwzwzs}
does not contribute to the quadratic growth. The effective
contribution of the heavy resonances remains.  The amplitude can be
approximated:
\begin{equation}
\label{e.pgwzwzs2} 
 \cm_{G^+ G^3 \to G^+ G^3} \approx { 1 - g_h^2 \over v^2}  t  \qquad m_h < E < m_1  
\end{equation}
If $g_h < 1$ the amplitude still grows quadratically, but slower
(unless we are in the Higgsless limit where $g_h=0$).  The quadratic
growth is further softened around the first resonance mass.  We can
approximate
\begin{eqnarray}
\label{e.pgwzwzs3} 
\nonumber \cm_{G^+ G^3 \to G^+ G^3} &\approx& 
 \left ({ 1 - g_h^2 \over v^2} - 3 {g_{C,1}^2 \over m_{C,1}^2} \right ) t   \\  
&-& g_{C,1}^2  \left ({t - s \over u - m_{C,1}^2} + {t - u \over s - m_{C,1}^2} \right )  
\qquad  E \sim m_1  
\end{eqnarray}
Above the first resonance the coefficient of the growing terms is
diminished by $3 g_1^2/m_1^2$.
 
In the model of Section \ref{sec:strongHolo} with the Higgs on the
brane we found $g_h = (v/\ti v)^3$.  This implies that $(1 -
g_h^2)/v^2 \approx 3/f_h^2$ in the Higgs limit, and $(1 - g_h^2)/v^2 =
1/v^2 \approx 1/f_h^2$ in the Higgsless limit.  Thus the growth of the
amplitude below the resonance scale is controlled by the scale $f_h$,
although the coefficient varies depending on which limit we consider.
In the model of Section \ref{s.hpgh} with a pseudo-goldstone Higgs
boson we found $g_h \approx \cos(\ti v/f_h)$.  This implies $(1 -
g_h^2)/v^2 \approx 1/f_h^2$ both in the Higgs limit and in the
Higgsless limit.

In both models, unitarity is restored by resonances at the scale
$\mkk$.  How efficiently the restoration proceeds, depends on the
coupling strength of the lowest lying resonance(s), which is clearly model
dependent. 
Nevertheless, in the 5D models we have studied we were
able to find useful approximations for these couplings. The results
for general backgrounds were given in Section \ref{s.cmcr} and Section
\ref{s.pgcr}.  In order to get some feeling about validity of our
estimates we now present the exact results for numerical calculations
in AdS$_5$.
 
We take the Planck/TeV hierarchy between UV and IR brane corresponding
to $a_L^{-1} = 10^{15}$.  The volume factor is then $\cv \approx k L
\approx 35$, the KK scale $\mkk \approx \pi k a_L$, the decay constant
$g_L f_h \approx 2 k a_L/\sqrt{\cv}$.

In the model of Section \ref{sec:strongHolo} the masses of lightest
resonances are found as
\begin{equation}
m_{W,1} \approx 0.77 \mkk \quad m_{W,2} \approx 0.78 \mkk
\end{equation} 
in the Higgs limit and 
\begin{equation}
m_{W,1} \approx 0.78 \mkk  \quad  m_{W,2} \approx 1.22 \mkk  
\end{equation} 
in the Higgsless limit.
In the Higgs limit we find the couplings $g_{W,1} \approx 8.2 g_W$, $g_{W,2} \approx 8.3 g_W$.
Thus, at the resonance scale there are two almost degenerate vector states with
comparable couplings to the electroweak Goldstone bosons.  These two
states saturate 66\% of the sum rule \erefn{cmsrh}.
In the Higgsless limit the couplings are $g_{W,1} \approx - 8.1 g_W$, $g_{W,2} \approx
0.1 g_W$, resulting in 96\% of the sum rule \erefn{cmsrhl} being saturated by the first resonance.
Thus the estimate below \eref{cmsrhl} perfectly captures the coupling of the first resonance. 
$W_2$ is the axial resonance and it approximately decouples from the electroweak Goldstone bosons. 
Avoiding violation of unitarity in the Higgsless limit requires $\mkk \simlt 1.5 \tev$ 
(that is $m_{W,1} \simlt 1.2 \tev$). 
Since $\mkk \approx \pi \sqrt{\cv} m_W \sim 1.5 \tev$, the bound is saturated with the current  choice of the hierarchy parameter $a_L^{-1} \sim 10^{15}$. 

In the model of Section \ref{s.hpgh} we find the masses of the
lightest resonances
\begin{equation}
m_{W,1} \approx 0.8 \mkk 
\quad 
m_{\bar W,1} \approx 0.8 \mkk
\quad  m_{W,2} \approx 1.2 \mkk  
\end{equation} 
where these masses are independent of the limit we consider.

In the Higgs limit we find $g_{W,1} \approx - g_{\bar W,1} \approx -
5.7 g_W$ while $g_{W,2} \approx 0$.  In the Higgsless limit the
coupling $g_{W,1} \approx - 8 g_W$, while $g_{\bar W,1} = 0$.  In both
cases, $95\%$ of the respective sum rule is saturated by the first
resonances.

The picture that emerges in both models is that the scattering
amplitude is almost entirely unitarized at the scale of the first
resonance.  A distinctive feature of the Higgs limit this is that
there are two almost degenerate resonances that contribute to
unitarisation, while in the Higgsless limit just one resonance does
most of the job.

\section{Conclusions}
\label{sec:conc} 
\setcounter{equation}{0}

New strong interactions as a cut-off to the SM are an interesting
alternative to supersymmetry as an explanation of the origin of the
electroweak scale. Before the LHC experiment tells us more about what
nature has chosen, it is important to investigate the possible ways
strong interactions could manifest themselves in the scattering of
longitudinally polarised vector bosons. This is necessary for staying
tuned in to the experimental analysis for various options.

We have studied models of electroweak breaking formulated in 5D warped space. 
By relying on the heuristic link to strongly interacting theories in 4D,
or simply referring to 5D models of electroweak symmetry breaking, we
have a technical, perturbative, means to investigate the detailed
dynamics of the longitudinally polarised electroweak gauge bosons.

In this paper we have studied the dynamics of gauge bosons in two,
previously proposed  5D models of electroweak symmetry breaking.
 Using the powerful mass eigenstate technique in  a general background \cite{defa,aa} we have calculated the mass spectrum
of the resonances and their couplings, as well as the couplings of the
physical `composite' Higgs boson to the Goldstone bosons eaten by the
$W$ and $Z$ gauge bosons. Thus we have obtained all the elements
necessary for calculating scattering amplitudes of the longitudinally
polarised vector bosons using the equivalence theorem. 
This allowed us to discuss the role of various contributions in unitarising
these scattering amplitudes.

Our explicit calculation in two very different models allows us to
extract quite general features, hopefully common to any strongly
interacting cut-off to the SM. These are, first of all, physical
scales that emerge and characterise the gauge boson dynamics. We have
identified four such scales: the electroweak scale $v$, the Higgs
boson decay constant or equivalently the Higgs boson composite scale
$f_h$, the resonance scale $M_{KK}$ and finally the cut-off scale
$\Lambda$ of the strongly interacting sector itself. Particularly
interesting are the relative values of the scale $v$ versus $f_h$, and
$f_h$ versus $M_{KK}$ which fully determine the unitarity violation and restoration in the gauge
bosons scattering amplitudes.

Another interesting common feature of the two models is that they
smoothly interpolate between the composite Higgs limit and the
Higgsless limit, depending on the relative magnitude of the scales
mentioned above. Thus we have explicit examples in which the strong
dynamics is not as simple as is usually assumed in Higgsless models
based on $SU(2)\times SU(2)\rightarrow SU(2)$ and the gauge boson
dynamics may lie between the composite Higgs and Higgsless limits. It
is interesting by itself that in the Higgsless limit we have
identified a new class of Higgsless models where the scalar particle
does not play any role in the unitarisation of the scattering
amplitudes but remains in the spectrum.

On the other side, our computations provide some insight into the phenomenology of gauge boson scattering. 
The picture that emerges from 5D models is  very different from the expectations based on simple unitarisation procedures of the effective chiral Lagrangians \cite{chunit}
For example, we do not have any scalar resonances in the spectrum. 
Moreover, the number and properties of the low-lying vector resonances are quite constrained. 

\section*{Acknowledgements}

We thank Piotr Chankowski for his contributions to this projects at its early stage. 
SP would like to thank Gian Giudice for useful discussions. 
The work of AF and SP was supported by the European Community Contract MRTN-CT-2004-503369 for the years 2004--2008 and by  the MEiN grant 1 P03B 099 29 for the years 2005--2007.
The work of JPR was funded under the FP6 Marie Curie contract MTKD-CT-2005-029466.

\renewcommand{\thesection}{Appendix \Alph{section}}
\renewcommand{\theequation}{\Alph{section}.\arabic{equation}} 
\setcounter{section}{0} 
\setcounter{equation}{0} 

\section{Warped Trigonometry} 
\label{a.wt} 

The equation of motion
\begin{equation}
\label{e.gme} 
\pa_5 (a^2(x_5)  \pa_5 f(x_5) ) + z^2 f(x_5) = 0 
\end{equation}
has two independent solutions.
Denote them $C(x_5,z)$ and $S(x_5,z)$. We choose them such their boundary conditions as 
\begin{equation}
C(0,z) = 1 \qquad C'(0,z) = 0 \qquad
S(0,z) = 0 \qquad S'(0,z) = z 
\end{equation}
so that in flat space they reduce to the familiar cosine and sine. 
The Wronskian relation 
\begin{equation} 
\label{e.wr}
S'(x_5,z) C(x_5,z) - C'(x_5,z) S(x_5,z) = z \, a^{-2}(x_5)
\end{equation} 
is the warped analog of $\sin^2 + \cos^2 = 1$. 

Integrating \eref{gme} twice, we  obtain the integral representation of the warped trigs 
\begin{eqnarray}
C(x_5,z) &=& 1 - z^2 \int_0^{x_5} a^{-2}(y)\int_0^{y} C(y',z)
\nn
S(x_5,z) &=& z \int_0^{x_5} a^{-2}(y)  - z^2 \int_0^{x_5} a^{-2}(y)\int_0^{y} S(y',z)
\end{eqnarray} 
from which follows the expansion at small $z$:
\begin{eqnarray}
\label{e.cssz} 
C(x_5,z) &=&  1 -  z^2 \int_0^{x_5}y  a^{-2}(y) 
+ z^4 \int_0^{x_5} a^{-2}(y)\int_0^{y} \int_0^{y'} y''  a^{-2}(y'') + \dots
\nn
S(x_5,z) &=&  z \int_0^{x_5} a^{-2}(y) 
- z^3 \int_0^{x_5} a^{-2}(y)\int_0^{y} \int_0^{y'} a^{-2}(y'') + \dots
\end{eqnarray}

For $z$ and $y$ such that $z^2/a^2 \gg 1 $ we have another useful approximation. 
\begin{eqnarray}
\label{e.wtwr} 
C(x_5,z) &\approx &      a^{-1/2}(x_5) \alpha \cos (z \int_L^{x_5}a^{-1}(y) + \phi_\alpha ) 
\nn
S(x_5,z) &\approx &   -   a^{-1/2}(x_5) \beta  \cos (z \int_L^{x_5}a^{-1}(y) + \phi_\beta)  
\end{eqnarray}
where the four real parameters  $\alpha$, $\beta$, $\phi_\alpha$, $\phi_\beta$ are bound to satisfy  
 $\beta \alpha \sin (\phi_\beta - \phi_\alpha) = 1$. 
Moreover, for metrics highly warped toward the IR brane  we have an approximate relation $C(L,z) \approx - z L S(L,z)$, $C'(L,z) \approx - z L S'(L,z)$ that follows from the perturbation expansion \erefn{cssz}, with $y \to L$ in the integrals. 

\vspace{1cm}

Let us see the warped sines and cosines in some particular, solvable backgrounds. 
 
For {\bf flat space}, $a(x_5) = 1$, we get the familiar trigonometric functions: 
\begin{equation}
C(x_5,z) = \cos (z x_5) \qquad   S(x_5,z) = \sin(z x_5)
\end{equation}

For AdS$_5$ we insert $a(x_5) = e^{-k x_5}$ and almost as easily  solve \eref{gme} in terms of Bessel functions. The solution is $a^{-1} Z_1(m/ka)$ 
(note also that $[a^{-1} Z_1(m/ka)]' = z a^{-2} Z_0(m/ka)$) and we pick up the following combinations 
\begin{eqnarray}
\label{e.csads}
C(x_5,z) &=& {\pi z \over 2 k} a^{-1}(x_5) \left [
 Y_0 \left ( {z \over k} \right )      J_1 \left ( {z \over k a(x_5)} \right ) 
- J_0 \left ( {z \over k} \right )     Y_1 \left ( {z \over k a(x_5)} \right )  
\right ]
\nn
S(x_5,z) &=&  {\pi z  \over 2 k} a^{-1}(x_5) \left [
- Y_1 \left ( {z \over k} \right )     J_1 \left ( {z \over k a(x_5)} \right )  
+ J_1 \left ( {z \over k } \right )      Y_1 \left ( {z \over k a(x_5)} \right ) 
\right ]
\end{eqnarray}

Another solvable background is that with a {power law} warp factor 
$a(x_5) = \left ( 1 - {k x_5 \over \gamma - 1} \right )^\gamma$
For $\gamma \to \infty$ we recover the exponential warp factor of the RS model.
The solutions to \eref{gme} can be written, similarly as in  the AdS$_5$ case, in terms of the Bessel functions, 
\begin{eqnarray}
C(x_5,z) & = &   {\pi z \over 2 k} a^{-1 + {1 \over 2\gamma} } \left [ 
Y_{1 \over 2\gamma-2}  \left ({z \over k } \right ) 
J_{2 \gamma-1 \over 2\gamma-2}  \left ({z \over k a^{1 - {1 \over \gamma}}} \right ) 
- J_{1 \over 2\gamma-2}  \left ({z \over k } \right ) 
Y_{2 \gamma-1 \over 2\gamma-2}  \left ({z \over k a^{1 - {1 \over \gamma}}} \right ) \right ]
\nn
S(x_5,z) & = & {\pi z \over 2 k}
a^{-1 + {1 \over 2\gamma}} \left [ 
- Y_{ 2 \gamma-1 \over 2\gamma-2}  \left ({z \over k } \right ) 
J_{2 \gamma-1 \over 2\gamma-2}  \left ({z \over k a^{1 - {1 \over \gamma}}} \right ) 
+ J_{2 \gamma-1 \over 2\gamma-2}  \left ({z \over k } \right ) 
Y_{2 \gamma-1 \over 2\gamma-2}  \left ({z \over k a^{1 - {1 \over \gamma}}} \right ) 
\right ]
\nn
\end{eqnarray}

\vspace{1cm}

\section{$SU(3)_c\times SU(2)_L\times SU(2)_R \times U(1)_X$: profiles and couplings}
\label{app:boundCond1}

We collect here various technical details concerning the model of Section 3.  
We list the profiles that follow from solving the boundary conditions on the IR brane. 
For the W tower
\bea
\label{Wtower1}
f_{L,n}^i  &=& \alpha_{W,n} C(x_5,m_{W,n})
\nn 
f_{R,n}^i  &=& - {g_R \over g_L}  \alpha_{W,n} { C'(L,m_{W,n}) \over S'(L,m_{W,n})}  S(x_5,m_{W,n})
\nn
(\alpha_{W,n})^{-2}  &=&
\int_0^L dy \left \{ C^2 (y,m_{W,n}) + {g_R^2 \over g_L^2} \left ({C'(L,m_{W,n}) \over S'(L,m_{W,n})} \right )^2  
S^2 (y,m_{W,n}) \right \} 
\eea
For the photon tower 
\bea 
f_{L,n}^3 &=& \sin \theta_W \alpha_{\gamma,n} C(x_5,m_n)
\nn
f_{R,n}^3 &=& s_x \cos \theta_W \alpha_{\gamma,n}C(x_5,m_n)
\nn
f_{X,n} &=& c_x \cos \theta_W \alpha_{\gamma,n} C(x_5,m_n)
\nn 
(\alpha_{\gamma,n})^{-2}  &=&  \int_0^L dy  C^2 (y,m_n)
\eea 
For the Z tower 
\bea 
f_{L,n}^3 &=& \cos \theta_W \alpha_{Z,n} C(x_5,m_{Z,n})
\nn
f_{R,n}^3 &=& -  s_x \sin \theta_W \alpha_{Z,n}  C(x_5,m_{Z,n})
 - {c_x^2 \over s_x} \sin \theta_W \alpha_{Z,n} { C'(L,m_{Z,n}) \over S'(L,m_{Z,n})} S(x_5,m_{Z,n})
\nn
f_{X,n} &=& -  c_x \sin \theta_W \alpha_{Z,n}  C(x_5,m_n)
 +  c_x \sin \theta_W \alpha_{Z,n} { C'(L,m_{Z,n}) \over S'(L,m_{Z,n})} S(x_5,m_{Z,n})
\nn
(\alpha_{Z,n})^{-2}  &=&
\int_0^L dy \left \{  C^2 (x_5,m_{Z,n}) 
+ {c_x^2 \over s_x^2} \sin^2 \theta_W  \left ({C'(L,m_{Z,n}) \over S'(L,m_{Z,n})} \right )^2 
S^2 (x_5,m_{Z,n}) \right \} 
\nn
\eea 
The quantization conditions are given in eqs. \erefn{cmwtq}, \erefn{cmptq}, \erefn{cmztq}. 
Expanding $C$ and $S$ at small $m$ allows to estimate the masses of W and Z. 
Including corrections of order $v^2/\mkk^2$ we find  
\bea&
\label{e.cwzmv} 
m_W^2 \approx g_L^2 {\ti v^2 \over 4} \left ( 
1 +  {\ti v^2 \over 4} \left [ 
 g_L^2 L^{-1} \int_0^L \int_0^y y' a^{-2}(y')  - g_L^2  \int_0^L y a^{-2}(y)    - g_R^2 L \int_0^L a^{-2}(y) \right ]\right)  
\nl
m_Z^2 \approx ( g_L^2+ g_Y^2) {\ti v^2 \over 4} \left ( 
1  \bnl
+  {\ti v^2 \over 4} \left [ 
 (g_L^2+g_Y^2) L^{-1} \int_0^L \int_0^y y' a^{-2}(y')  - g_L^2  \int_0^L y a^{-2}(y)    - g_R^2 L \int_0^L a^{-2}(y)
+  g_Y^2\int_0^L\int_0^y a^{-2}(y')   \right ]  \right)  
\nn
\eea
From dimensional analysis one would expect the integrals in the above expression to be of order $1/\mkk^2$.  
However, some of the integrals scale linearly with $L$, therefore they can be enhanced when the volume factor is large.  
In such a case, the corrections turn out to be $\co(v^2/f_h^2)$ rather than $\co(v^2/\mkk^2)$.  

The Goldstone profiles corresponding to W and Z are given by
\bea
\label{e.cmegp}
\bar f_{L,W}^i  &=& \alpha_{W} m_W^{-1} C'(x_5,m_W)
\nn 
\bar f_{R,W}^i  &=& - {g_R \over g_L}  \alpha_{W} { C'(L,m_W) \over S'(L,m_W)}  m_W^{-1}  S'(x_5,m_W)
\nn
\ti f_W^i  &= & - \alpha_{W} {2 \over m_{W} \sqrt L  g_L \ti v } a_L^2 C'(L,m_{W})
\eea 
\bea 
\bar f_{L,Z}^3 &=& \cos \theta_W \alpha_{Z} m_Z^{-1} C'(x_5,m_Z)
\nn
\bar f_{R,Z}^3 &=& -  s_x \sin \theta_W \alpha_{Z}  m_Z^{-1}  C'(x_5,m_Z)
 - {c_x^2 \over s_x} \sin \theta_W \alpha_{Z} { C'(L,m_Z) \over S'(L,m_Z)} m_Z^{-1}  S'(x_5,m_Z)
\nn
\bar f_{X,Z} &=& -  c_x \sin \theta_W \alpha_{Z}  m_Z^{-1}  C'(x_5,m_Z)
 +  c_x \sin \theta_W \alpha_{Z} { C'(L,m_Z) \over S'(L,m_Z)} m_Z^{-1}  S'(x_5,m_Z)
\nn
\ti f_Z^3  &= &
- \alpha_{Z} {2 \over m_{Z} \sqrt L  \sqrt{g_L^2 + g_Y^2} \ti v } a_L^2 C'(L,m_{Z})
\eea 
The expansion of these profiles for small masses  can be done with the help of \eref{cssz}. 
To lowest order in $m_W$, $m_Z$, we derive  \eref{cmwsis}.

The profiles serve to establish the couplings of the Goldstones to the resonances. 
Working out the relevant terms in the 5D action  the couplings to the neutral gauge bosons are given by  
\bea &
\label{e.cmrc} 
g_{N,n} = {m_W^2 \over \sqrt L} \int_0^L dy a^{-2}(y) \left \{ 
  g_L  y^2 f_{L,n}^3(y)  
+ {g_R^3 \over g_L^2}  L^2 f_{R,n}^3(y)
 \right \}
\nl 
+ {1 \over 2} \sqrt{L} {v^2 \over \ti v^2} (g_L f_{L,n}^3(L) + g_R f_{R,n}^3(L))  
\eea
while those to the charged gauge bosons 
\bea &
g_{C,n} =  {m_W^2 \over \sqrt L} \int_0^L dy a^{-2}(y) \left \{ 
  g_L  y^2 f_{L,n}^i(y)  
+ {g_R^3 \over g_L^2}  L (s_x^2 y + c_x^2 L)  f_{R,n}^i(y)
 \right \}
\nl 
+ {1 \over 2} \sqrt{L} {v^2 \over \ti v^2} (g_L f_{L,n}^i(L) + g_R f_{R,n}^i(L))  
\eea
Due to the $a^{-2}$ factor, the integrals are dominated by the behaviour of the profiles near the IR brane. Therefore it is sane to replace $y \to L$ under the integrals. 
Moreover, we set $g_L = g_R$ for simplicity. 
This yields   
\begin{eqnarray} 
\label{e.cmrcwzg} 
g_{W,n} &\approx&  g_L \sqrt{L} \int_0^L a^{-2}(y) 
\left \{  {1 \over 2} {v^2 \over \ti v^2} \delta(L) +  m_W^2  L a^{-2}(y) \right \}
\nn&& 
{ 
C(y,m_{W,n}) - {C'(L,m_{W,n}) \over S'(L,m_{W,n})} S(y,m_{W,n})
\over 
\left [\int_0^L dy \left \{ 
C^2(y,m_n) 
+ \left ({C'(L,m_n) \over S'(L,m_n)} \right )^2   S^2 (y,m_n) \right \} \right]^{1/2}} 
\nn
g_{Z,n} &\approx&  {g_L^2 - g_Y^2 \over \sqrt{g_L^2 + g_Y^2}} \sqrt{L} \int_0^L a^{-2}(y) 
\left \{  {1 \over 2} {v^2 \over \ti v^2} \delta(L) +  m_W^2  L a^{-2}(y) \right \}
\nn && 
{C(y,m_{Z,n}) - {C'(L,m_{Z,n}) \over S'(L,m_{Z,n})} S(y,m_{Z,n}) 
\over 
\left[ \int_0^L dy \left \{  C^2 (y,m_n) 
+ \cos 2\theta_W  \left ({C'(L,m_n) \over S'(L,m_n)} \right )^2 
S^2 (y,m_n) \right \} \right]^{1/2}
}  
\nn
g_{\gamma,n} &\approx&  2 e \sqrt{L} \int_0^L a^{-2}(y) 
\left \{  {1 \over 2} {v^2 \over \ti v^2} \delta(L) +  m_W^2  L a^{-2}(y) \right \}
{ C(y,m_{\gamma,n}) \over 
\left[\int_0^L C^2 (y,m_{\gamma,n})
 \right ]^{1/2}  
 }
\end{eqnarray}

\section{$SU(3)_c\times SO(5) \times U(1)_X$: profiles and couplings}
\label{app:boundCond}

We move to the holographic model of a pseudo-Goldstone Higgs boson. 
Choosing the direction of the Higgs vev as  
$\la  A_5^{\hat 4}   \ra  = \ti v  a^{-2}/\sqrt {\int_0^L a^{-2}(y)}$,  
the Wilson-line matrix is given by 
\begin{equation}
\omega(x_5,\ti v)  = \left [ \ba{ccccc}
 1& 0& 0& 0& 0\\
 0& 1& 0& 0& 0\\ 
 0& 0& 1& 0& 0\\ 
 0& 0& 0& \cos \left ( {\ti v \over f(x_5)} \right )&-  \sin \left ( {\ti v \over  f(x_5)} \right )\\ 
 0& 0& 0& \sin \left ( {\ti v \over  f(x_5)} \right ) & \cos \left ( {\ti v \over  f(x_5)} \right )
\ea 
\right ] \quad
f(x_5) = {\rt \sqrt{\int_0^L a^{-2}(y)} \over g \sqrt{L} \int_0^{x_5} a^{-2}(y) } 
\end{equation}    
This yields the link between the profiles with zero and  non-zero vev:  
\begin{eqnarray}
\label{e.wt}
f_L^a(x_5,\ti v) &=& {1 + \cos(\ti v/f) \over 2} f_L^a(x_5)
+ {1 - \cos(\ti v/f) \over 2} f_R^a(x_5) 
+ {\sin(\ti v/f) \over \rt} f_C^a(x_5)
\nn
f_R^a(x_5, \ti v) &=& {1 - \cos(\ti v/f) \over 2} f_L^a(x_5)
+ {1 + \cos(\ti v/f) \over 2} f_R^a(x_5) 
- {\sin(\ti v/f) \over \rt} f_C^a(x_5)
\nn
f_C^a(x_5, \ti v) &=& - {\sin(\ti v/f) \over \rt}  f_L^a(x_5)
+ {\sin(\ti v/f) \over \rt}  f_R^a(x_5) 
+ \cos(\ti v/f) f_C^a(x_5)
\nn
f_C^4(x_5, \ti v) &=& f_C^4(x_5) \qquad \qquad  f_X(\ti v) = f_X(x_5)
\end{eqnarray}
Inserting this into the IR boundary conditions we can find the mass eigenstates. 

\bc \bf Charged \ec 

The charged gauge bosons are combinations of $L_\mu^i$,$R_\mu^i$ and
$C_\mu^i$.  In the charged sector we have two towers.  In the one
referred to as the {\bf $\bar W$ tower} the masses do not depend on
$\ti v$.  The quantization condition is simple:
\begin{equation}
S'(L,m_{\bar W,n}) = 0  
\end{equation}  
This tower has the profiles:
\begin{eqnarray} 
f_{R,n}^i &=& \alpha_{\bar W,n} \sqrt{2} \cos(\ti v/ f_h)
S(x_5,m_{\bar W,n}) 
\nn f_{C,n}^i &=& - \alpha_{\bar W,n} \sin(\ti v/f_h) S(x_5,m_{\bar W,n})
\nn
(\alpha_{\bar W, n} )^{-2} &=& (1 + \cos^2(\ti v/f_h))^2 \int_0^L
[S(y,m_{\bar W,n})]^2
\eea

In the other tower, referred to as the {\bf $W$ tower}, the masses do
depend on $\ti v$.  The quantization condition is
\begin{equation}
\label{e.wtqc}
C'(L,m_{W,n}) S(L,m_{W,n}) + {1 \over 2} m_{W,n} a_L^{-2} \sin^2\left
( \ti v \over f_h\right) = 0
\end{equation}  
The corresponding profiles are 
\begin{eqnarray} 
f_{L,n}^{i} &=& \alpha_{W,n} C(x_5,m_{W,n}) \nn 
f_{R,n}^i &=& -
\alpha_{W,n} {C'(L,m_{W,n}) \over S'(L,m_{W,n})} S(x_5,m_{W,n}) \nn
f_{C,n}^i &=& - \alpha_{W,n} {\rt \cos(\ti v/ f_h) \over \sin(\ti
v/f_h)} {C'(L,m_{W,n}) \over S'(L,m_{W,n})} S(x_5,m_{W,n})
\nn
(\alpha_{W,n})^{-2} &=& \int_0^L \left \{ 
[C(y,m_{W,n})]^2
- {C'(L,m_{W,n}) C(L,m_{W,n})\over S'(L,m_{W,n}) S(L,m_{W,n})} [S(y,m_{W,n})]^2
\right \}
\eea
There is a light solution of the quantization condition proportional
to $f_h \sin(\ti v/f_h)$. This is the $W$ boson.

\bc \bf Neutral \ec 

The neutral gauge bosons are combinations of $L_\mu^3$,$R_\mu^3$ and
$C_\mu^{3,4}$ and $X_\mu$.  
Three of them have $\ti v$-independent masses. 
One is along the same group space direction as the Higgs vev, hence we refer to it as the {\bf Higgs tower}.  
The quantization condition:
\begin{equation}
S(L,m_{H,n}) = 0 
\end{equation}  
The profile 
\begin{equation}
f_{C,n}^4 = \alpha_{H,n} S(x_5 ,m_{H,n})
\qquad
(\alpha_{H,n})^{-2} = \int_0^L [S(y,m_{H,n})]^2 
\end{equation}
There is no light (mode) in this tower. 

Another is called the {\bf photon tower}.  
The quantization condition:
\begin{equation}
C'(L,m_{\gamma,n}) = 0  
\end{equation}
The profiles 
\begin{eqnarray}
f_{L,n}^3 &=& \sin \theta_W \alpha_{\gamma,n}  C(x_5,m_{\gamma,n})
\nn
f_{R,n}^3 &=&  s_x \cos \theta_W \alpha_{\gamma,n} C(x_5,m_{\gamma,n})
\nn
f_{X,n} &=&  c_x \cos \theta_W \alpha_{\gamma,n}  C(x_5,m_{\gamma,n})
\nn
(\alpha_{\gamma,n})^{-2} &=& \int_0^L [C(y,m_{\gamma,n})]^2
\eea
The photon tower includes a massless eigenvector: the photon.

There is the {\bf $\bar Z$ tower}, which is similar to the $\bar W$
tower and has no light mode.  The quantization:
\begin{equation}
S'(L,m_{\bar Z,n}) = 0
\end{equation}
The profiles 
\begin{eqnarray}
f_{R,n}^3 &=& - c_x \alpha_{\bar Z,n} \sqrt{2} \cos(\ti v/ f_h)
S(x_5,m_{\bar Z,n}) \nn f_{X,n} &=& s_x \alpha_{\bar Z,n} \sqrt{2}
\cos(\ti v/ f_h) S(x_5,m_{\bar Z,n}) \nn f_{C,n}^3 &=& c_x
\alpha_{\bar Z,n} \sin(\ti v/ f_h) S(x_5,m_{\bar Z,n})
\nn
(\alpha_{\bar Z, n} )^{-2} &=& (1 + \cos^2(\ti v/f_h) - s_x^2 \sin^2(\ti v/f_h))^2 \int_0^L [S(y,m_{\bar Z,n})]^2
\eea

Finally there is the {\bf $Z$ tower}, where masses are sensitive to $\ti
v$.  The quantization condition is very similar to that of the W-tower,
\begin{equation}
\label{e.ztqc}
\cos^2 \theta_W C'(L,m_{Z,n}) S(L,m_{Z,n}) + {1 \over 2} m_{Z,n}
a_L^{-2} \sin^2\left(\ti v \over f_h\right) = 0
\end{equation} 
the difference being the cosine of the Weinberg angle.
The profiles  
\begin{eqnarray}
f_{L,n}^3 &=& \cos \theta_W \alpha_{Z,n} C(x_5,m_{Z,n}) \nn f_{R,n}^3
&=& - s_x \sin \theta_W \alpha_{Z,n} C(x_5,m_{Z,n}) - c_x^2
\cos\theta_W \alpha_{Z,n} {C'(L,m_{Z,n}) \over S'(L,m_{Z,n})}
S(x_5,m_{Z,n}) \nn f_{X,n} &=& - c_x \sin \theta_W \alpha_{Z,n}
C(x_5,m_{Z,n}) + c_x \sin\theta_W \alpha_{Z,n} {C'(L,m_{Z,n}) \over
S'(L,m_{Z,n})} S(x_5,m_{Z,n}) \nn f_{C,n}^3 &=& - \cos\theta_W
\alpha_{Z,n} {\rt \cos( \ti v/f_h) \over \sin(\ti v/f_h)}
{C'(L,m_{Z,n}) \over S'(L,m_{Z,n})} S(x_5,m_{Z,n})
\nn
(\alpha_{Z,n})^{-2} &=&   \int_0^L \left \{ 
[C(y,m_{Z,n})]^2
- {C'(L,m_{Z,n}) C(L,m_{Z,n})\over S'(L,m_{Z,n}) S(L,m_{Z,n})} [S(y,m_{Z,n})]^2
\right \}
\eea
The $Z$ boson is the lightest solution of the quantization condition
with the mass proportional to $f_h \sin(\ti v /f_h)$.

We move to discussing the profiles of the Goldstones corresponding to
the electroweak gauge bosons.  The Goldstone profiles at zero vev are
simply related to the corresponding gauge profiles, see \eref{gfh}.
The Goldstones eaten by $W$ and $Z$ have the following profile
\begin{eqnarray} 
\label{e.pggpe} 
\bar f_{L,W}^i &=&   \alpha_{W}  m_W^{-1} C'(x_5,m_{W}) 
\nn
\bar f_{R,W}^i &=&  -  \alpha_{W} {C'(L,m_{W}) \over S'(L,m_{W})}  m_W^{-1} S'(x_5,m_{W})
\nn
\bar f_{C,W}^i &=& 
 -  \alpha_{W}  {\rt \cos(\ti v/ f_h) \over  \sin(\ti v/f_h)} {C'(L,m_{W}) \over S'(L,m_{W})} m_W^{-1} S'(x_5,m_{W}) 
\nn
\bar f_{L,Z}^3 &=& \cos \theta_W \alpha_{Z} m_Z^{-1} C'(x_5,m_{Z})
\nn
\bar f_{R,Z}^3 &=& -  s_x \sin \theta_W \alpha_{Z} m_Z^{-1} C'(x_5,m_{Z}) 
-  c_x^2 \cos\theta_W \alpha_{Z,n}  {C'(L,m_{Z}) \over S'(L,m_{Z})} m_Z^{-1} S'(x_5,m_{Z})
\nn
\bar f_{X,Z} &=& - c_x \sin \theta_W \alpha_{Z}  m_Z^{-1}  C'(x_5,m_{Z,n})
+   c_x \sin\theta_W \alpha_{Z}  {C'(L,m_{Z}) \over S'(L,m_{Z})} m_Z^{-1}  S'(x_5,m_{Z})
\nn
\bar f_{C,Z}^3 &=&
 - \cos\theta_W \alpha_{Z}  {\rt \cos( \ti v/f_h) \over  \sin(\ti v/f_h)} {C'(L,m_{Z}) \over S'(L,m_{Z})} S'(x_5,m_{Z}) 
\end{eqnarray} 
Expanding the warped trigs for a small  $m$ yields the approximate expressions   \erefn{hpgwsis}. 

These approximate profiles allow us to determine the resonance couplings. 
The general formula for the charged ones is  
\begin{eqnarray} & 
g_{C,n} = g \sqrt{L} \int_0^L a^2 \left \{ f_{L,n}^i \bar f_{L,Z}^3 \bar
f_{L,W}^i + f_{R,n}^i \bar f_{R,Z}^3 \bar f_{R,W}^i + {1 \over 2}
\left ( f_{L,n}^i + f_{R,n}^i \right ) \bar f_{C,Z}^3 \bar f_{C,W}^i
\bnl + {1 \over 2} f_{C,n}^i \left ( \bar f_{L,Z}^3 + \bar f_{R,Z}^3
\right )\bar f_{C,W}^i - {1 \over 2} f_{C,n}^i \left ( \bar f_{L,W}^i
+ \bar f_{R,W}^i \right )\bar f_{C,Z}^3 \right\}
\end{eqnarray} 
(no summing over $i$ here).  For the neutral ones,
\begin{eqnarray} & 
g_{N,n} = g \sqrt{L} \int_0^L a^2 \left \{ f_{L,n}^3 \bar f_{L,W}^i \bar
f_{L,W}^i + f_{R,n}^3 \bar f_{R,W}^i \bar f_{R,W}^i + {1 \over 2}
\left ( f_{L,n}^3 + f_{R,n}^3 \right ) \bar f_{C,W}^i \bar f_{C,W}^i
\bnl f_{C,n}^3 \left ( \bar f_{L,W}^i + \bar f_{R,W}^i \right )\bar
f_{C,W}^i \right\}
\end{eqnarray} 

Inserting the approximate Goldstone profiles and the exact profiles of
vector resonances we obtain the couplings
\begin{equation}
g_{W,n} \approx {g_L \over 2} {\sqrt{L} \int_0^L a^{-2}(y) \left \{
C(y,m_{W,n}) - {C'(L,m_{W,n})\over S'(L,m_{W,n})} S(y,m_{W,n}) ( 1 -
s_x^2 (1-y/L)) \right \} \over \left [ \int_0^L a^{-2}(y) \right ]
\left [ \int_0^L C^2(y,m_{W,n}) - {C'(L,m_{W,n}) C(L,m_{W,n})\over
S'(L,m_{W,n}) S(L,m_{W,n})} S^2(y,m_{W,n})\right ]^{1/2} }
\end{equation}
\begin{equation}
g_{\hat W,n} \approx {g_L \over \rt} {\cos(\ti v/f_h) \over \sqrt{1 +
\cos^2(\ti v/f_h)}} {\sqrt{L} \int_0^L a^{-2}(y) S(y,m_{\hat W,n})
\over \left [ \int_0^L a^{-2}(y)\right ] \left [\int_0^L
S^{2}(y,m_{\hat W,n}) \right ]^{1/2} }
\end{equation} 
\begin{equation}
g_{Z,n} \approx {g_L^2-g_Y^2 \over 2 \sqrt{g_L^2 + g_Y^2}} {\sqrt{L}
\int_0^L a^{-2}(y) \left \{ C(y,m_{Z,n}) - {C'(L,m_{Z,n})\over
S'(L,m_{Z,n})} S(y,m_{Z,n}) (1 - {2 g_L^2 \cos^2(\ti v/f_h) \over
g_L^2 - g_Y^2} (1 - y/L) ) \right \} \over \left [ \int_0^L
a^{-2}(y)\right ] \left [ \int_0^L C^2(y,m_{Z,n}) - {C'(L,m_{Z,n})
C(L,m_{Z,n})\over S'(L,m_{Z,n}) S(L,m_{Z,n})} S^2(y,m_{Z,n})\right
]^{1/2} }
\end{equation}
\begin{equation}
g_{\bar Z,n} \approx {g_L \over \rt} c_x {\cos(\ti v/f_h) \over \sqrt{1 +
\cos^2(\ti v/f_h) - s_x^2 \sin^2(\ti v/f_h)}} {\sqrt{L} \int_0^L
a^{-2}(y) S(y,m_{\hat Z,n}) \over \left [ \int_0^L a^{-2}(y)\right ]
\left [\int_0^L S^{2}(y,m_{\hat Z,n}) \right ]^{1/2} }
\end{equation} 
\begin{equation}
g_{\gamma,n} \approx {e} {\sqrt{L} \int_0^L a^{-2}(y) C(y,m_{\gamma,n})
\over \left [ \int_0^L a^{-2}(y)\right ] \left [\int_0^L
C^{2}(y,m_{\gamma,n}) \right ]^{1/2} }
\end{equation} 
The Higgs tower does not couple to the electroweak Goldstones.


\end{document}